\documentclass[sigconf]{acmart}
\usepackage{multirow}
\usepackage{booktabs}
\usepackage{enumitem} 

\AtBeginDocument{%
  }

\copyrightyear{2026}
\acmYear{2026}
\setcopyright{cc}
\setcctype{by}
\acmConference[CHI '26]{Proceedings of the 2026 CHI Conference on Human Factors in Computing Systems}{April 13--17, 2026}{Barcelona, Spain}
\acmBooktitle{Proceedings of the 2026 CHI Conference on Human Factors in Computing Systems (CHI '26), April 13--17, 2026, Barcelona, Spain}
\acmDOI{10.1145/3772318.3790873}
\acmISBN{979-8-4007-2278-3/2026/04}

\begin{document}

\title{Growing With the Condition: Co-Designing Pediatric Technologies that Adapt Across Developmental Stages}


\author{Neda Barbazi}
\affiliation{%
  \institution{College of Design}
  \institution{University of Minnesota}
  \city{Minneapolis, Minnesota}
  \country{USA}}
\email{barba087@umn.edu}

\author{Ji Youn Shin}
\affiliation{%
  \institution{College of Design}
  \institution{University of Minnesota}
  \city{Minneapolis, Minnesota}
  \country{USA}}
\email{shinjy@umn.edu}

\author{Gurumurthy Hiremath}
\affiliation{%
  \institution{Medical School}
  \institution{University of Minnesota}
  \city{Minneapolis, Minnesota}
  \country{USA}}
\email{hiremath@umn.edu}

\author{Carlye Anne Lauff}
\affiliation{%
  \institution{College of Design}
  \institution{University of Minnesota}
  \city{Minneapolis, Minnesota}
  \country{USA}}
\email{carlye@umn.edu}

\renewcommand{\shortauthors}{Barbazi et al.}

\begin{abstract}
  Children with chronic conditions face evolving challenges in daily activities, peer relationships, and clinical care. Younger children often rely on parental support, while older ones seek independence. Prior studies on chronic conditions explored proxy-based, family-centered, and playful approaches to support children’s health, but most technologies treat children as a homogeneous group rather than adapting to their developmental differences. To address this gap, we conducted four co-design workshops with 69 children with congenital heart disease (CHD) at a medically supported camp, spanning elementary, middle, and high school groups. Our analysis reveals distinct coping strategies: elementary children relied on comfort objects and reassurance, middle schoolers used mediated communication and selective disclosure, and high schoolers emphasized agency and direct engagement with peers and providers. Through child-centered participatory design, we contribute empirical insights into how children’s management of chronic conditions evolves and propose design implications for pediatric health technologies that adapt across developmental trajectories.
\end{abstract}




\maketitle

\section{Introduction}
Over one in four children worldwide lives with chronic conditions such as diabetes, autism, or congenital heart disease (CHD), which reshape daily routines and developmental trajectories \cite{wisk2025prevalence,mastorci2025chronic, barbazi_understanding_2025, barbazi_perceiving_2023}. These long-term conditions impact children and families, requiring ongoing medical attention and continual adaptation across developmental stages \cite{mastorci2025chronic,mastorci2024transition,oster2024chdpulse}. Beyond medical care, chronic illness disrupts daily activities, schooling, peer relationships, and psychosocial well-being \cite{mastorci2025chronic,stinson2016ipeer2peer}, often restricting participation \cite{alexandridis2021rubys,hong2020diaries}, generating stigma \cite{dunbar2024journey, su2024stigma}, and undermining identity \cite{su2024cysticfibrosis,liu2015friends}. Children strive to feel “normal” while balancing treatment regimens and activity limitations \cite{ankrah2022worlds,alexandridis2021rubys,liu2015friends}. 

Among chronic illnesses, CHD is the most common congenital condition, affecting about 1\% of newborns worldwide and nearly 40,000 annually in the United States \cite{cdc2024chd}. Unlike heart conditions that develop later in life, CHD results from structural differences that form during pregnancy and are present from birth. CHD was once frequently fatal in early childhood, but medical advances now enable children to survive into adolescence and adulthood \cite{burns2022chdliteracy, oster2024chdpulse}. However, because long-term survival is relatively recent, many children lack the developmental and psychosocial supports available for other chronic conditions \cite{ang2023qualitycare, mastorci2025chronic, barbazi2025scoping, barbazi2025assessments}. Living with CHD often involves early surgeries, visible scars or implanted devices, and regular medical visits that shape daily routines in distinct ways \cite{oster2024chdpulse, arya2013parentsCHD}. For example, a typical day may involve balancing activity limitations at school, handling classmates’ questions about scars, and managing medications/frequent check-ups \cite{burns2022chdliteracy, rodts2020literacy}. These experiences also shift as children grow.

Across developmental stages, children’s self-management strategies often become insufficient, requiring families to renegotiate routines and caregiving roles \cite{boris2024mentalhealth,gordonjames2023ai, seo2019caregiving, seo2021parentchild}. Younger children rely on parents for reassurance, advocacy, and help with routines \cite{su2024stigma}, while adolescents take on responsibilities related to privacy, disclosure, and independence in decision-making \cite{seo2025chatbot,zehrung2024transitioning,seo2021parentchild,arya2013parentsCHD}. These transitions highlight the need for pediatric health technologies that can support children across early, middle, and late childhood rather than target a single age group \cite{cha2025healthtracking,cha2023diabetes,su2024systematicreview,lehnert2022cci}. Frequent switching between technologies introduces friction, disrupts routines, and burdens families managing complex care \cite{pina2017family}. Previous HCI research on pediatric chronic illness care examined multiple approaches to supporting children and families. Early systems positioned parents as proxies, focusing on monitoring tools, data logs, and communication platforms \cite{seo2021parentchild,pina2017family,kaziunas2015transition}. Family-centered designs expanded the scope to siblings, teachers, and peers, recognizing illness management as a relational ecosystem \cite{cha2025healthtracking,Nikkhah2022FamilyCare}. More recent child-centered designs foregrounded children through playful methods such as gamification, storytelling agents, and socially assistive robots to reduce anxiety, improve literacy, and sustain engagement \cite{zhou2024lollipop,isbister2021robot,jeong2018huggable, saksono2020storywell, barbazi_understanding_2025, zeng_octos_2025}. This progression reflects a shift from parent-focused to family-centered to child-centered designs  \cite{druin1999cooperative,guha2013ci}, yet most systems still assume a stable developmental stage rather than children’s changing needs.

Within this shift, co-design approaches position children as active contributors to design \cite{buckmayer2024pdchildren,qi2025participatory,sevon2023youngchildren,banker_usability_2022}, revealing insights that extend beyond technical features. In diabetes care, \citet{cha2023diabetes} co-designed a diabetes management game to support the transition from childhood to adolescence, revealing everyday strategies such as “educated guessing." \citet{foster2023robot} showed that children’s input on robot roles and behaviors is essential for reducing anxiety and pain, highlighting that relational qualities matter as much as technical ones. \citet{hijab2025bagofstuff} used multi-sensory co-design with autistic and non-autistic children to integrate diverse sensory and communication needs into play tools. Narrative and fictional scaffolding further support children in articulating sensitive socio-emotional experiences \cite{warren_lessons_2022}, while validated child personas ground design in lived experiences \cite{warnestal_effects_2017}. Warren et al. \cite{warren_codesign_2023} similarly argue that digital health systems must adapt as children’s needs evolve.

Despite these advances, three gaps remain. First, many studies treat children as a homogeneous group or substitute parents’ perspectives for children’s voices \cite{islind2025proxy,hamidi2017proxies}. Second, even when children participate directly, most research focuses on a single developmental stage, overlooking how needs shift as children grow \cite{su2024systematicreview,ahmadpour2023trauma,koutna2021growth}. Third, very few studies compare children’s perspectives across age groups within the same chronic condition, limiting understanding of developmental transitions in underserved populations like CHD. To address these gaps, we conducted four co-design workshops with children with CHD at a camp, involving elementary, middle, and high school groups. Using age-appropriate participatory methods, we elicited and compared their challenges and coping strategies across age groups. Two research questions (RQs) guided our study: RQ1: What challenges do children with CHD identify across different developmental stages, and how do their coping strategies shift across these stages? RQ2: What design implications emerge for supporting developmentally adaptive pediatric health interventions? From these questions, our contributions are twofold: \enlargethispage*{16pt}

\begin{itemize}
    \item We provide empirical insights into the lived experiences of children with CHD, highlighting the challenges they face and how coping strategies shift across developmental stages.
    
    \item We identify design implications for creating developmentally adaptive pediatric health interventions that evolve alongside children as they grow.
     
\end{itemize}

These contributions advance pediatric HCI by centering developmental differences within CHD and demonstrating how cross-age co-design can inform adaptive health technologies in chronic care. In the following sections, we present related work, methods, findings, and discussion.

\section{Related Work}
In this section, we review HCI research on pediatric chronic care and co-design with children. We highlight technologies for care management, participatory approaches across domains, and limitations in chronic pediatric care.

\subsection{Technologies for Pediatric Chronic Illness Management} 
HCI researchers have long examined how chronic illness management extends beyond the clinic into everyday family life, particularly in pediatrics, where children lack full capacity for self-care. Early studies characterized parents as proxies, responsible for translating medical instructions into daily routines, monitoring symptoms, and coordinating communication with providers \cite{su2024cysticfibrosis,seo2021parentchild,pina2017family,kaziunas2015transition}. This framing emphasized caregiving labor as a central form of technological engagement, as parents relied on tracking systems, monitoring apps, or paper artifacts to sustain adherence. More recent scholarship reframes this role by describing caregivers as boundary actors who bridge medical and family domains, negotiating tensions between clinical protocols and everyday parenting needs \cite{richards2025collectiveroutines,bhat2023familycaregivers,Nikkhah2022FamilyCare}. This perspective shows how technologies not only facilitate information transfer but also shape routines and trade-offs in caregiving, as seen in systems that support parental decision-making and mediate family–provider communication \cite{seo2025chatbot,seo2019caregiving}. 

As this perspective broadened, research began to situate pediatric chronic illness within distributed ecologies of care, where responsibility extends beyond parents to siblings, teachers, peers, and extended family. Studies of cancer, cystic fibrosis, diabetes, and asthma highlight how schools and households coordinate health routines, producing care ecologies—networks of people, artifacts, and practices that collectively sustain children’s well-being across daily contexts \cite{cha2025healthtracking,su2024cysticfibrosis,cha2023diabetes,sonney2022asthmaapp,seo2021parentchild}. Technological interventions range from coordination platforms that link hospitals, homes, and schools \cite{Nikkhah2022FamilyCare} to systems that support “collective routines” across households \cite{richards2025collectiveroutines}. These works establish that pediatric illness care is not confined to the immediate family; it requires tools that can support fluid collaboration across multiple caregivers and contexts. Yet much of this research continued to position children as recipients of care rather than contributors to its design \cite{cha2025healthtracking,chow2024familycentred, shin_design_2018, shin_towards_2019}.

In response, a parallel stream of work explored playful and educational technologies that actively scaffold children’s engagement with chronic illness, shifting attention to active participation. Serious games, interactive storybooks, and social robots have been deployed to reduce anxiety, teach medical concepts, and sustain motivation for treatment adherence \cite{sarasmita2024games,zhou2024lollipop,isbister2021robot,jeong2018huggable}. Probe-based interventions such as DreamCatcher \cite{pina2020dreamcatcher} and Snack Buddy \cite{schaefbauer_snack_2015} demonstrate how tracking can become a collaborative, family-centered activity, while exergames like Storywell \cite{saksono2020storywell} show how shared play fosters both child participation and caregiver motivation. Reviews confirm gamification remains a dominant strategy for pediatric engagement \cite{sarasmita2024games, stutvoet_gamification_2024}. However, while these solutions effectively scaffold interaction and emotional engagement, they typically assume a single developmental stage and emphasize short-term participation rather than long-term developmental change that characterizes chronic conditions.

More recently, the field has shifted toward adaptivity, designing technologies that dynamically adjust to developmental needs or health states. For example, biofeedback-based interventions in therapeutic contexts demonstrate how task difficulty can be adjusted in real time to sustain engagement \cite{antunes2023digitaltwin,frutos2014adhd}, whereas adaptive eHealth platforms tailor feedback or learning materials to children’s cognitive capacities \cite{su2024systematicreview,wang2024adaptiveui}. Adaptability also applies at the family level, where shifting caregiving roles require flexible technological support \cite{nikkhah2024resilience}. Adolescents’ use of diabetes and asthma management apps reveals how adolescents negotiate tensions between independence, privacy, and parental oversight, underscoring the importance of tools that adjust as children mature \cite{wyche2025diabetes,zehrung2024transitioning}. Warren et al. \cite{warren_codesign_2023} similarly show how children’s and families’ strategies for navigating digital health systems change across developmental stages and contextual transitions—arguing for platforms that support “shifting needs” rather than fixed-use patterns. In chronic conditions, such adaptability is especially critical, as children’s developmental stages influence how they communicate, interpret, and engage with digital health tools \cite{warnestal_effects_2017}.
Pediatric HCI has begun acknowledging developmental change, but most technologies still target a single age group and rarely consider how children’s capacities shift across early childhood, middle childhood, and adolescence. While this growing body of research demonstrates increasing attention to adaptation, few studies have examined how the design process itself might support such responsiveness. Because developmental changes shape how children interpret and manage their condition, understanding adaptability requires engaging them at different stages of growth. This need underscores the value of co-design approaches that account for children’s developmental differences and involve them directly in shaping technologies.\enlargethispage*{16pt}

\subsection{Co-design with Children}
Co-design has become central in pediatric HCI because it positions children as active partners rather than passive recipients of care. Foundational child–computer interaction work introduced methods such as Cooperative Inquiry, which builds intergenerational design teams through techniques like Mixing Ideas (children and adults sketch and merge concepts) and Layered Elaboration (overlays that support iteration without erasing contributions) \cite{druin1999cooperative,guha2004mixing,walsh2010layered}. Later methods broadened participation to younger and more diverse groups. The Mosaic Approach provides preschool children with cameras, drawings, and mapping activities to express experiences beyond literacy \cite{clark2011mosaic}, while Bags of Stuff uses tangible materials to let children “show” their ideas \cite{yip2013brownies}. More recently, the Collaborative Design Thinking (CoDeT) framework systematized co-design in high child-to-adult ratio settings by assigning structured peer roles to support equitable collaboration \cite{vanmechelen2019codet}. Across these methods, aligning activities with children’s developmental abilities and communication preferences enables researchers to surface tacit knowledge, emotions, and experiential insights that inform design.

In pediatric healthcare, co-design supports children’s participation in their own care while making medical routines less intimidating and more meaningful. Approaches span playful systems that turn treatment into games, family-centered apps that coordinate illness management, and sensory-sensitive environments for neurodiverse children. Unlike other domains, pediatric healthcare introduces constraints—ethical concerns, clinical risks, and family dynamics—that shape how participation can occur. Sometimes children participate directly: \citet{vandelden2020spiroplay} co-designed SpiroPlay, a breathing game that transformed spirometry into playful at-home practice, reducing stress and encouraging routine rehearsal. When direct involvement is not possible, proxies may represent children’s perspectives. \citet{hamidi2017proxies} introduced Participatory Design with Proxies (PDwP), engaging parents, teachers, and therapists to speak for children with limited communication. Recent pediatric HCI research also underscores the need for adaptive and developmentally sensitive co-design; for example, \citet{warnestal_effects_2017} showed how co-designed child personas embed developmental insights, and \citet{warren_lessons_2022} demonstrated how narrative scaffolding supports age-appropriate discussion of sensitive experiences. These studies reflect a shift from parent-proxy approaches to family-centered care, to child-centered participation, and more recently to developmentally adaptive pediatric technologies. Yet most technologies and co-design studies still focus on a single developmental stage, limiting opportunities to compare how children’s perspectives and coping strategies change over time.

Co-design with children in other domains faces some similar constraints but also offers methodological practices less common in pediatric care. In education, co-design supports surfacing children’s values around fairness, autonomy, and collaboration through the design of classroom technologies, interactive narratives, and educational robots \cite{obaid2024classroomrobots,liu2025narrative,alvesoliveira2021robotdesigners,alvesoliveira2017yolo}. As in healthcare, children shape both technical and relational qualities, yet educational settings more often employ structured multimodal and narrative scaffolds—collaborative storytelling, character-driven ideation, and embodied role play—that help children externalize abstract ideas and emotions. For example, \citet{alvesoliveira2021robotdesigners} showed that children shaped an educational robot’s personality and interaction style through character-driven storytelling. More broadly, socio-emotional design work demonstrated how narrative and participatory storytelling methods help children articulate conflict, empathy, and emotional reasoning in developmentally meaningful ways \cite{zhou2024lollipop}. These approaches appear more selectively in pediatric care, where clinical constraints limit the full use of narrative and expressive methods. Together, these disciplinary differences suggest that supporting children’s emotional expression depends not only on the activities offered but on how co-design participation is structured and scaffolded.

This developmental sensitivity becomes especially visible in how co-design is structured. Across domains, studies vary widely in duration and developmental focus. Many are short, one-off workshops with no follow-up \cite{obaid2024classroomrobots,buckmayer2024pdchildren,frauenberger2017blending}, while others run across several sessions tailored to children’s attention and abilities \cite{zhou2024lollipop}. A smaller number extend over multiple years; for example, \citet{alvesoliveira2021robotdesigners} conducted a two-year robot design project that introduced new cohorts at each stage to support design iteration rather than track developmental change. In healthcare, co-design typically centers on condition-specific contexts tied to narrow developmental windows and shaped by ethical and clinical constraints \cite{chow2024familycentred,sarasmita2024games}. These structural choices matter because children’s ability to express needs, emotions, and coping strategies is closely linked to developmental capacities such as attention, verbal reasoning, emotional regulation, and prior illness experience. Short sessions often surface only immediate preferences, whereas multi-session or age-specific designs reveal deeper developmental differences in how children interpret challenges and participate in co-design. What becomes visible therefore depends strongly on how participation is organized—by age group, session length, cohort structure, or methodology \cite{buckmayer2024pdchildren,sevon2023youngchildren}—underscoring that developmental needs involve not only the activities used but the structure and pacing of participation over time.

Prior research in pediatric HCI shows a shift toward involving children more directly in their care, yet co-design studies rarely account for developmental transitions or compare experiences across age groups. Despite progress, several gaps remain. First, although many studies focus on condition-specific technologies, we know little about how children articulate challenges and coping strategies at different developmental stages, making it difficult to understand how needs shift with age. Second, most pediatric interventions are designed for a single age group and tied to a specific developmental window, meaning children often “age out” as they move from parent-dependent care toward greater autonomy. Third, co-design studies themselves are frequently shaped by practical constraints—such as one-off workshops or narrow age cohorts—that prioritize feasibility over capturing developmental change. As a result, both co-design processes and the technologies they generate rarely account for how children’s roles and needs shift over time, and solutions are seldom evaluated longitudinally or designed to adapt across these transitions. Addressing these gaps requires co-design that adapts to different age ranges, captures developmental differences within a chronic condition, and generates design implications that evolve as children grow.\enlargethispage*{16pt}

\section{Methods}

We designed a child-centered co-design workshop to elicit age-specific challenges and coping strategies from children with chronic conditions and to compare these across developmental stages, addressing RQ1.

\subsection{Study Context, Settings and Participants}
\begin{figure}[h]
  \centering
  \includegraphics[width=\linewidth]{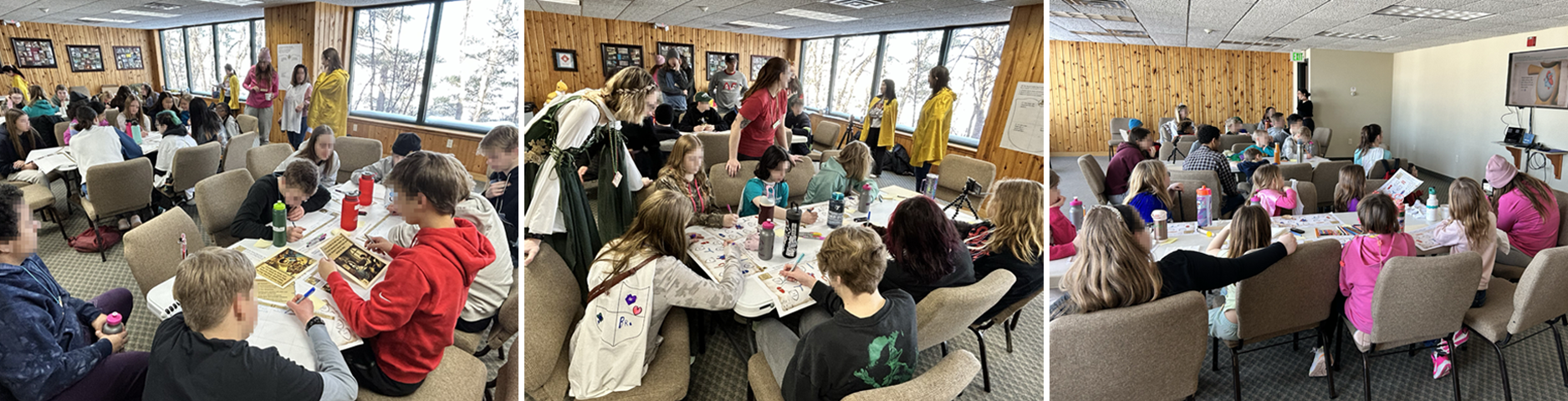}
  \caption{Co-design workshop sessions with high school (left), middle school (center), and elementary school (right) groups.}
  \Description{Fig 1. Three-panel figure showing three sessions of a co-design workshop. The left photo shows high school students sitting around three tables, using paper forms and markers, while five adult facilitators stand in the room to help children. The center photo shows middle school students sitting at three tables, using paper forms and markers, while nine adult facilitators stand nearby, helping children. The right photo shows elementary school students sitting around three tables, using paper forms and markers, while watching a movie on a TV on the wall, with an adult facilitator.}
  \label{fig:Fig1}
\end{figure}

Our study involved children diagnosed with congenital heart disease (CHD), whose ongoing care across childhood and adolescence offers a lens on how their  challenges and coping strategies evolve as they grow. Existing interventions often center on parents and providers, leaving children’s perspectives underrepresented \cite{rodts2020literacy,arya2013parentsCHD}. Yet children actively make sense of their condition and develop their own coping strategies. Capturing these insights is essential for designing tools that adapt to developmental change \cite{oster2024chdpulse,ang2023qualitycare,burns2022chdliteracy,rodts2020literacy,arya2013parentsCHD}. To address this gap, we conducted a qualitative, age-comparative co-design workshop with elementary, middle, and high schoolers in four back-to-back 45-minute sessions during a three-day, medically supported, community-based Camp Odayin’s winter camp \cite{CampOdayin_About} for children with CHD. This camp setting provided consistent conditions and facilitation for comparing developmental differences across groups. The program combines recreational activities with continuous medical oversight, creating a safe, inclusive environment where children share a diagnosis and peer community. Unlike clinics that evoke medical anxiety or school settings that can introduce stigma, the camp offered a neutral, comfortable space that supported open participation. It also made it feasible to engage multiple age groups simultaneously under the same medical supervision, enabling systematic comparison across developmental stages without adding stressors or disrupting care (see Figure~\ref{fig:Fig1}).

Our workshop was embedded within the camp’s daily schedule as a one-day session in February 2025, held from morning to noon, to explore how children (1) make sense of their condition, (2) communicate lived experiences, and (3) articulate support needs. Each session aligned with the camp’s activity blocks to minimize fatigue and maintain routine continuity. We developed an ongoing collaboration with the camp over several months through regular communication and prior visits, which established trust between our research team and camp staff. All participants were pre-registered camp attendees diagnosed with CHD. The camp conducts its own admission and medical screening to ensure children are prepared for recreational and group activities; we did not recruit independently. Instead, families learned about the study through our partnership with the camp and opted in if they met eligibility criteria: (1) a confirmed CHD diagnosis; (2) age between roughly 5 and 18 years; (3) physical and cognitive ability to participate in light group activities; and (4) English proficiency sufficient to follow prompts and express ideas verbally or through drawing. Children with medical or emotional instability that could interfere with participation were excluded. Families provided written parental permission under an IRB approval from the University of Minnesota (STUDY00020670), and each child gave verbal assent at the start of their session. Participation was voluntary, and children could leave or skip activities at any time. No identifying or sensitive health data were collected beyond eligibility verification, and all data were anonymized.

A total of 69 children participated in four sessions organized by developmental stage: high school (HS, n=20; 29.0\%), middle school (MS, n=33; 47.8\% across two sessions), and elementary school (ES, n=16; 23.2\%). Together, the groups included 36 boys (52.2\%) and 33 girls (47.8\%). Within each group, HS consisted of 14 girls (70.0\%) and 6 boys (30.0\%); MS included 12 girls (36.4\%) and 21 boys (63.6\%); and ES comprised 7 girls (43.8\%) and 9 boys (56.2\%). Participants remained in their pre-assigned camp cohorts, with each cohort seated across 3–4 tables in a multipurpose room. HS cohorts included Cohort 1 (n=6), Cohort 2 (n=8), and Cohort 3 (n=6). MS cohorts included Cohort 4 (n=7), Cohort 5 (n=6), Cohort 6 (n=7), Cohort 7 (n=6), and Cohort 8 (n=7). ES cohorts included Cohort 9 (n=7), Cohort 10 (n=6), and Cohort 11 (n=3). Demographic info on race or socioeconomic background was not collected due to camp’s policy (see Table~\ref{tab:participants}). 

\begin{table*}[t]
\centering
\caption{Participant distribution by session, educational stage, and cohort (n=69).}
\resizebox{\textwidth}{!}{%
\setlength{\tabcolsep}{15pt}
\begin{tabular}{lllllll}
\toprule
Session & Educational Stage & Cohort & n & \% of total & Girls (n, \%) & Boys (n, \%) \\
\midrule
\multirow{3}{*}{Session 1} 
 & \multirow{3}{*}{\shortstack{High School (HS) \\ ~14--18 years}} 
 & Cohort 1 & 6 & 8.7\% & 6 (100.0\%) & 0 (0.0\%) \\
 & & Cohort 2  & 8 & 11.6\% & 8 (100.0\%) & 0 (0.0\%) \\
 & & Cohort 3   & 6 & 8.7\%  & 0 (0.0\%)   & 6 (100.0\%) \\
\textbf{Subtotal HS} & & & \textbf{20} & \textbf{29.0\%} & \textbf{14 (70.0\%)} & \textbf{6 (30.0\%)} \\
\midrule
\multirow{2}{*}{Session 2} 
 & \multirow{2}{*}{\shortstack{Middle School (MS) \\ ~13--14 years}} 
 & Cohort 4 & 7 & 10.1\% & 0 (0.0\%) & 7 (100.0\%) \\
 & & Cohort 5     & 6 & 8.7\%  & 6 (100.0\%) & 0 (0.0\%) \\
\textbf{Subtotal MS} & & & \textbf{13} & \textbf{18.8\%} & \textbf{6 (46.2\%)} & \textbf{7 (53.8\%)} \\
\midrule
\multirow{3}{*}{Session 3} 
 & \multirow{3}{*}{\shortstack{Middle School (MS) \\ ~11--12 years}} 
 & Cohort 6 & 7 & 10.1\% & 0 (0.0\%) & 7 (100.0\%) \\
 & & Cohort 7    & 6 & 8.7\%  & 6 (100.0\%) & 0 (0.0\%) \\
 & & Cohort 8   & 7 & 10.1\% & 0 (0.0\%) & 7 (100.0\%) \\
\textbf{Subtotal MS} & & & \textbf{20} & \textbf{29.0\%} & \textbf{7 (35.0\%)} & \textbf{14 (65.0\%)} \\
\midrule
\multirow{3}{*}{Session 4} 
 & \multirow{3}{*}{\shortstack{Elementary School (ES) \\ ~5--11 years}} 
 & Cohort 9 & 7 & 10.1\% & 7 (100.0\%) & 0 (0.0\%) \\
 & & Cohort 10 & 6 & 8.7\%  & 0 (0.0\%) & 6 (100.0\%) \\
 & & Cohort 11 & 3 & 4.3\%  & 0 (0.0\%) & 3 (100.0\%) \\
\textbf{Subtotal ES} & & & \textbf{16} & \textbf{23.2\%} & \textbf{7 (43.8\%)} & \textbf{9 (56.2\%)} \\
\midrule
\textbf{Total} & & & \textbf{69} & \textbf{100\%} & \textbf{33 (47.8\%)} & \textbf{36 (52.2\%)} \\
\bottomrule
\end{tabular}}
\label{tab:participants}
\end{table*}

A multidisciplinary team facilitated the workshops. The first author led all sessions, drawing on over two years of experience engaging children with CHD in clinical and community settings, including multiple prior visits to the camp. The team also included two senior co-authors specializing in child-centered health design, and an undergraduate design assistant. A pediatric cardiologist co-author advised on clinical accuracy, developmental sensitivity, and the framing of questions about challenges and lived experiences. Preparation began months in advance and involved co-creation with camp staff. Camp staff—including counselors and program directors with extensive experience supporting children with CHD—reviewed and refined all facilitation materials. Their feedback led to adjustments such as clarifying prompts, avoiding potentially triggering terminology, and balancing individual and group work. They also recommended warm-up activities and guidance on integrating digital and physical materials. All materials were aligned with the camp’s Medieval Renaissance theme to support continuity within the program. During each session, the first author guided the overall flow, narrated activities, and managed transitions, while senior researchers and the undergraduate assistant facilitated discussions at the tables. Each cohort was accompanied by two rotating camp counselors, a pediatric nurse, and a leadership staff member who ensured emotional safety, clarified children’s verbal or drawn ideas when needed, and monitored well-being throughout the activities. This structure created a developmentally responsive, medically supported environment and the trust needed for sensitive co-design work.\enlargethispage*{16pt}

\subsection{Design of the Co-Design Process}
\begin{figure*}[t]
  \centering
  \includegraphics[width=0.8\linewidth]{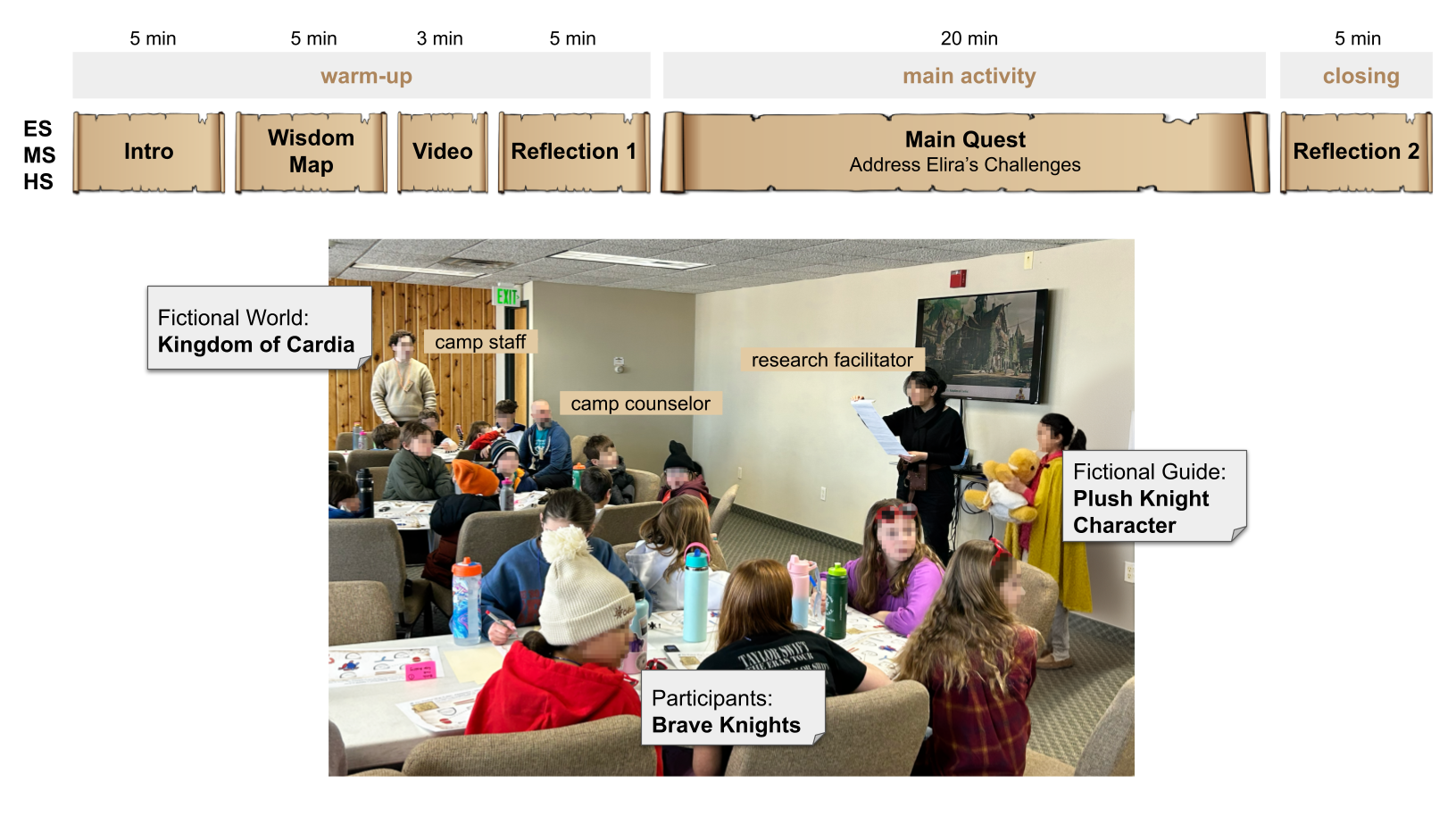}
  \caption{Study design and activity timeline (top), introduction scene with the “plush knight” as fictional guide, participants as “brave knights,” and camp and research facilitators in the Kingdom of Cardia (bottom).}
  \Description{Three-panel figure showing the study design, activity timeline, and the introduction scene to the Kingdom of Cardia. The top panel shows the warm-up, main activity, and closing sequence for ES, MS, and HS groups. The middle panel displays the activity flow. The bottom panel shows an introduction scene in which a research facilitator reads the Kingdom of Cardia story on behalf of the plush knight character and introduces Lady Elira, a child with a special heart, while another facilitator holds the plush knight. Children, participating as “brave knights,” sit across several tables with two camp staff and counselors nearby.}
  \label{fig:Fig2}
\end{figure*}

We grounded the workshop in participatory design (PD), which emphasizes mutual learning, situated engagement, and participant agency \cite{bodker2022pd,simonsen2013handbook,druin1999cooperative}. Mutual learning supported shared exploration, situated engagement ensured activities occurred in a familiar and safe context, and agency enabled children to contribute in age-appropriate ways. To adapt PD to a medically sensitive context, we used fictional inquiry, a narrative method that creates symbolic distance through story worlds where children can safely explore real experiences through metaphor \cite{dindler2007fictional,dindler2005mars,hiniker2017fictional}. Fictional inquiry helps children express thoughts and feelings that may be difficult to share directly \cite{warren_lessons_2022}. To align with the camp’s Medieval Renaissance theme, we situated the workshop in a fictional world called the \textit{Kingdom of Cardia} (a nod to cardiology). Children entered as “brave knights” on a quest, guided by a "plush knight character" \cite{barbazi2025boundary} that appeared in the room and across all workshop materials. Because the knight could not speak, the facilitator read an old-style letter on its behalf. The letter welcomed children to Cardia, explained that some villagers are born with special hearts, and introduced \textit{Lady Elira} (a title fitting the fictional era), a seven-year-old villager who feels unsure about her condition. Her story centered on three challenges: keeping up with friends during play, explaining her heart to others, and managing fear during visits to the healer (doctor). The plush knight then invited participants to help Lady Elira by completing each activity. This narrative provided symbolic distance, supported psychological safety, and offered continuity across the session (see Figure~\ref{fig:Fig2}).

\begin{table*}[t]
\centering
\renewcommand{\arraystretch}{1.2} 
\caption{Sequential activities in each co-design session with children.}
\resizebox{\textwidth}{!}{%
\begin{tabular}{p{1cm} p{3cm} p{9cm} p{2cm}}
\toprule
\textbf{Step} & \textbf{Activity} & \textbf{Purpose} & \textbf{Format} \\
\midrule
1 & Wisdom Map & Assess children’s heart knowledge and act as an icebreaker. & Individual \\
2 & Educational Video & Introduce key heart anatomy and prepare children for the Main Quest. & Group \\
3 & Reflection 1 & Reflect on feelings about the video. & Group \\
4 & Main Quest & Tackle 3 challenges of Lady Elira through age-appropriate activities:
\begin{itemize}[leftmargin=1.2em, itemsep=0pt, topsep=0pt, partopsep=0pt, parsep=0pt]
  \item drawing (elementary)
  \item mix of drawing and writing (middle school)
  \item writing (high school)
\end{itemize}& Individual\\[-11pt]
5 & Reflection 2 & Reflect on the co-design session’s feelings. & Group \\
\bottomrule
\end{tabular}}
\label{tab:activities}
\end{table*}

Over two years, the research team collaborated with pediatric cardiology providers (cardiologists, child-life specialists, nurses, and camp educators) and conducted interviews, surveys, and observations with children, parents, and providers in both clinic and camp settings. Together with CHD educational literature review \cite{barbazi2025scoping} and earlier phases of our broader study \cite{barbazi2025boundary, barbazi2025assessments, zeng_octos_2025}, these activities informed and validated the content of Lady Elira’s three narrative challenges through expert-informed review, ensuring clinical grounding and developmental appropriateness. This framing positioned children as players in a shared narrative and experts in their own experiences. We intentionally structured the workshop as a warm-up sequence that built children’s confidence and readiness so they felt prepared for the Main Quest, where they designed solutions for Lady Elira’s challenges using age-appropriate formats (see Table~\ref{tab:activities}).\enlargethispage*{16pt}

Each session followed five sequential activities using both hands-on and digital formats across individual and group work: (1) the Wisdom Map to assess heart anatomy knowledge and (2) an educational video introducing basic anatomy. Together, these activities supported ‘mutual learning’ by helping the research team see what children already knew while letting children learn something new so they felt they were gaining, not just giving. (3) The emotion-reflection then helped children talk about feelings in a low-pressure way and eased the transition to the Main Quest. We designed these warm-ups to reduce anxiety and avoid shaping children’s final responses, since they focused on basic anatomy while the Main Quest focused on challenges, coping, and support. (4) In the Main Quest, developmentally tailored scaffolds supported ‘agency’: younger children used drawing, stickers, and verbal explanations (with facilitators writing down their words), while older children used layered prompts that encouraged analysis and solution-building. (5) A short closing reflection helped children feel proud of their contributions. Embedding the activities in the camp and aligning them with its theme created ‘situated engagement’, and the story of Elira supported psychological safety by allowing children to explore challenges of belonging, disclosure, and clinical encounters without direct self-disclosure.\enlargethispage*{16pt}

\subsubsection{Icebreaker and Assessment: Wisdom Map (5 minutes)}
\begin{figure}[h]
  \centering
  \includegraphics[width=\linewidth]{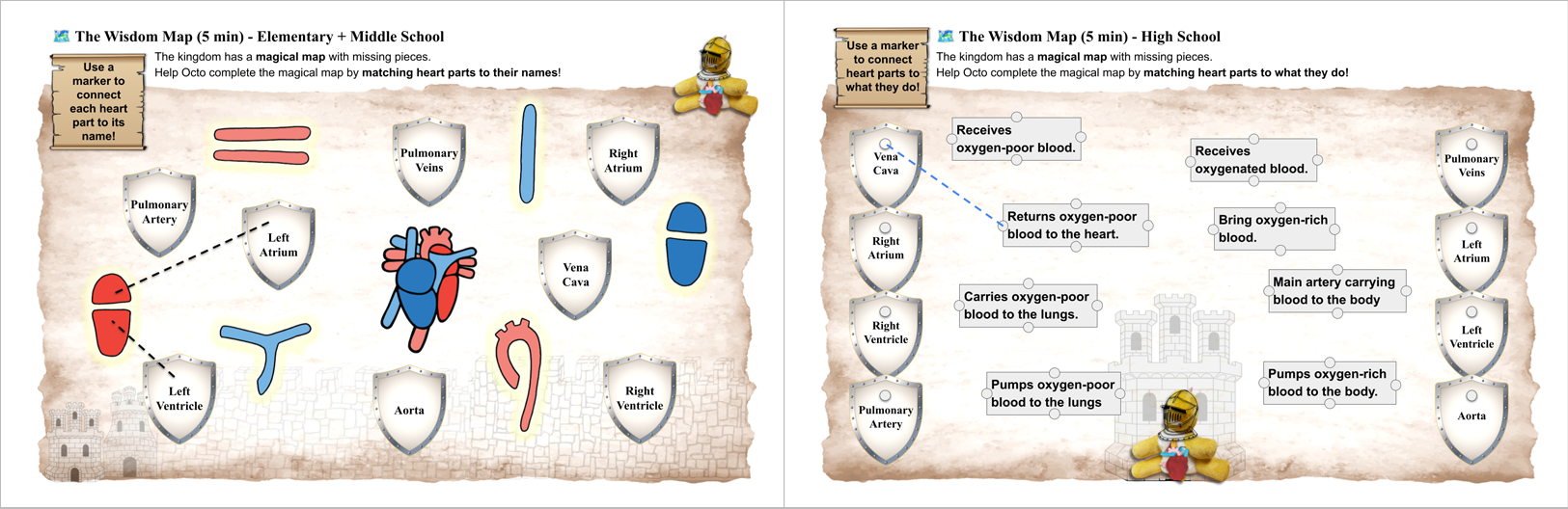}
  \caption{Wisdom Map worksheets for elementary and middle school children (left) and high school children (right).}
  \Description{Two-panel figure showing the Wisdom Map game-like matching activity. The left panel displays the worksheet for elementary and middle school groups, where children match visual heart parts with simple labels. The right panel shows the high school worksheet, where terms can be linked to short functional descriptions.}
  \label{fig:Fig3}
\end{figure}

The workshop began with the Wisdom Map, a game-like matching activity that introduced heart anatomy and positioned children as knowledgeable contributors. Each child received a parchment-style worksheet with heart parts scattered like puzzle pieces. Their task was to connect the pieces to names or functions, depending on the age group. Elementary and middle school participants matched visual cues with simple labels (e.g., “pulmonary artery,” “left atrium”), while high school participants linked terms to short functional descriptions (e.g., “carries oxygen-poor blood to the lungs”) (see Figure~\ref{fig:Fig3}). A five-minute timer, medieval-style music, and a projected village scene transformed the task into a playful challenge as an icebreaker. Although the maps were designed to be self-explanatory, facilitators and counselors provided light support as needed. The activity served two purposes: (1) to provide an informal baseline of knowledge, giving the research team a developmental snapshot; and (2) to build children’s confidence by allowing them to feel like knowledgeable contributors, reinforcing PD’s emphasis on mutual learning and agency \cite{sevon2023youngchildren,frauenberger2017blending,walsh2010layered,guha2008specialneeds}. At the end, children received a Wisdom Map Key showing both names and functions of the heart’s parts. While not central to data collection, this key served as a take-home learning aid and a small token of accomplishment.

\subsubsection{Educational Short Video (3 minutes)}
After the Wisdom Map, children watched a 2.5-minute animated video that explained heart anatomy through a train metaphor. The video had a main character that guided viewers along the “tracks” carrying oxygen-rich and oxygen-poor blood through the body. Unlike the individual, paper-based Wisdom Map, the video created a shared moment of digital attention. This shift balanced physical and digital media, moving children from working independently to learning together. The video served three purposes. (1) Reinforce anatomical content and build confidence for children uncertain after the Wisdom Map (several revisited worksheets to refine answers); (2) provide a shared foundation across age groups, ensuring participants entered the design phase with a common language; (3) encourage collective learning to support PD principles of mutual learning and equitable participation \cite{bannon2012designmatters,bodker2022pd}.

\subsubsection{Reflection Activity 1 (5 minutes)} 
\begin{figure}[h]
  \centering
  \includegraphics[width=0.88\linewidth]{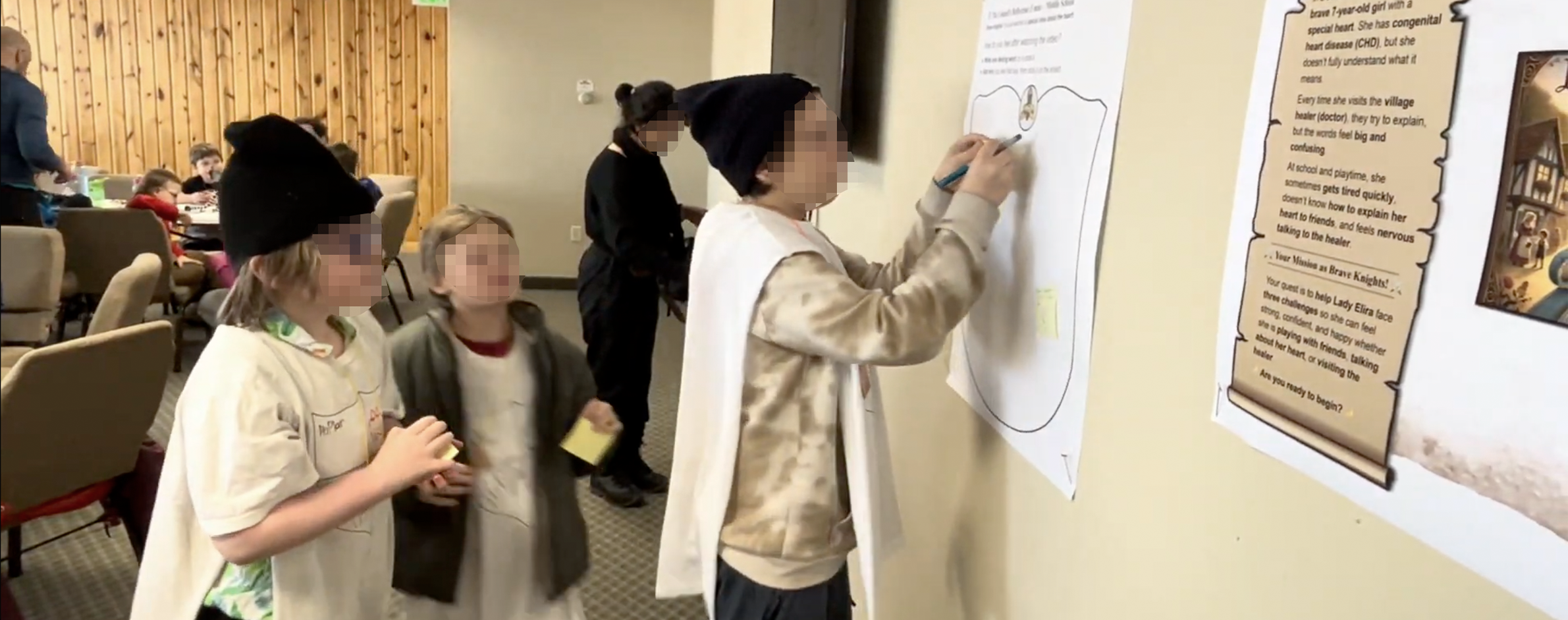}
  \caption{Reflection 1: Middle school children posting their reflections about the educational video.}
  \Description{Three middle schoolers are completing the first reflection after watching the educational video by posting sticky notes on large, shield-shaped posters attached to the wall, with an adult facilitator nearby. A camp counselor and additional children seated at tables are visible in the background.}
  \label{fig:Fig4}
\end{figure}

Immediately after the video, children completed a brief reflection using sticky notes on large, shield-shaped posters (see Figure~\ref{fig:Fig4}). Prompts were adapted by age: elementary participants wrote one feeling word, while middle and high school participants added a short explanation of their feeling. Facilitators and counselors supported participation by reading prompts aloud, helping with writing, and encouraging verbal contributions when needed. This simple, tactile activity created a shared moment of reflection that transitioned the group from passive watching to active engagement. It provided a safe outlet for emotional expression and reinforced PD principles of agency and voice by making children’s perspectives visible to peers \cite{sevon2023youngchildren,vanmechelen2019codet,walsh2010layered,yip2013brownies}. It also bridged the collective focus of the video with the imaginative problem-solving of the main design quest, ensuring children entered the next stage emotionally engaged and ready to contribute.

\subsubsection{Main Design Quest: Lady Elira’s Challenges (20 minutes)}
\begin{figure}[h]
  \centering
  \includegraphics[width=0.9\linewidth]{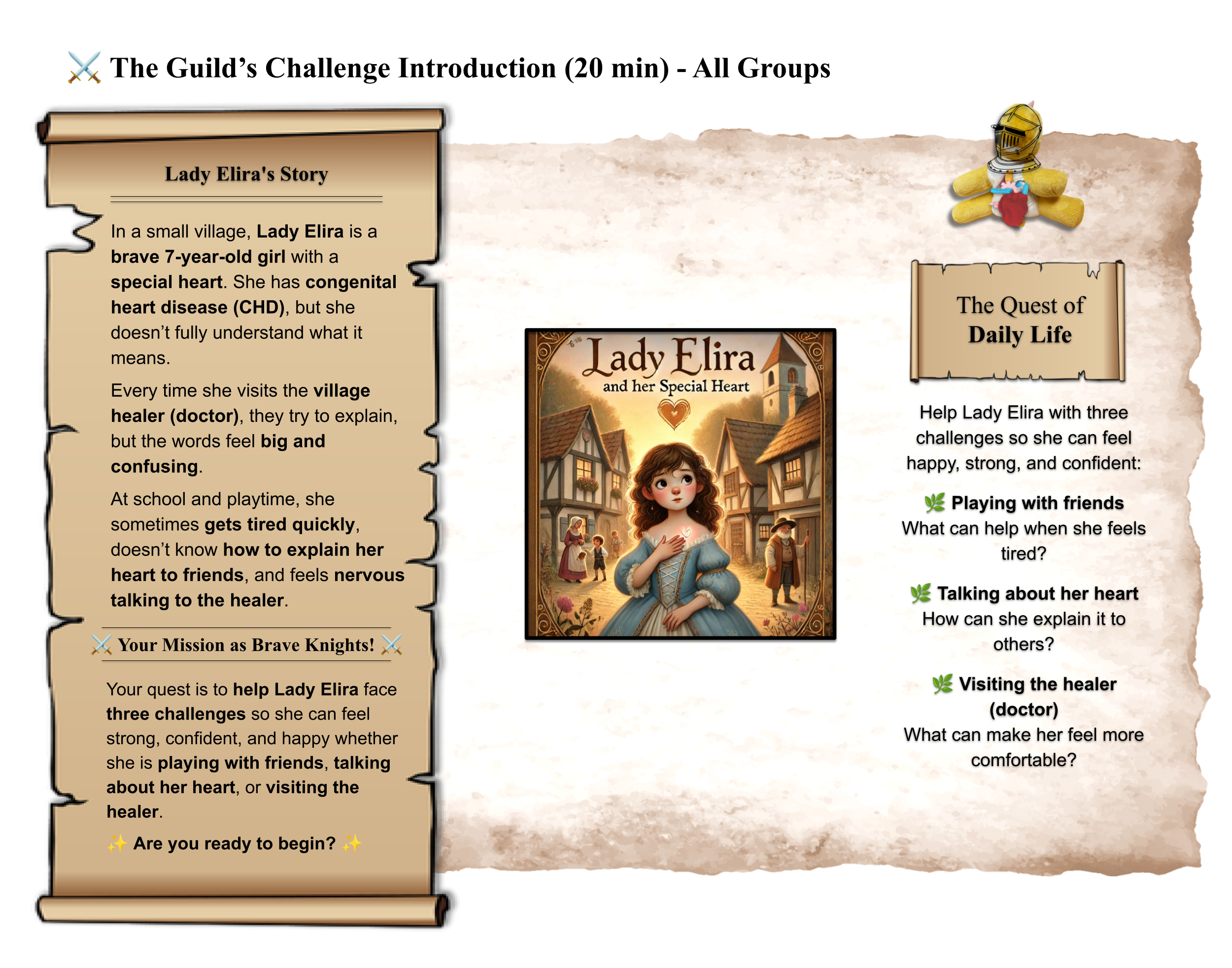}
  \caption{Main Design Quest persona poster introducing Lady Elira and her three challenges.}
  \Description{Main Design Quest persona poster introduces Lady Elira, a seven-year-old girl with a congenital heart condition in a medieval cottage scene. It provides a short story about her daily life and highlights three main challenges she faces: feeling tired while playing with friends, not knowing how to explain her heart condition to others, and feeling nervous about visiting the healer.}
  \label{fig:Fig5}
\end{figure}

The core activity centered on Lady Elira, a fictional peer with a “special heart,” whose story framed the children’s design contributions. The quest began with a live storytelling segment delivered beside a poster-sized persona introducing Elira’s three challenges: (1) fatigue during play, (2) difficulty explaining her condition, and (3) anxiety about visiting the healer (doctor) (see Figure~\ref{fig:Fig5}). 

\begin{table*}[t]
\centering
\renewcommand{\arraystretch}{1} 
\caption{Worksheet formats for the Main Design Quest.}
\resizebox{\textwidth}{!}{%
\begin{tabular}{p{3cm} p{4cm} p{6cm} p{3cm}}
\toprule
\textbf{Age Group} & \textbf{Worksheet Structure} & \textbf{Prompts} & \textbf{Mode of Expression} \\
\midrule
Elementary School &
Drawing-based worksheet with 3 large illustrated panels (one per challenge). &
\textit{“Show how she feels.” “Show what helps her feel better.” “Show how she can tell her friends she needs a break.” “Show how she explains her special heart and CHD.” “Show who/what helps her feel strong when she explains her special heart.” “Show how she feels about visit.” “Show how she can explain her feeling to doctor.” “Show who/what helps her feel safe and calm at the healer’s cottage.”} &
Drawing, stickers, minimal writing and verbatim notes \\
\midrule
Middle School &
3 × 3 table (rows = 3 challenges, columns = Problem, Tips for Elira, Tips for Others). &
\textit{“Why do you think this is hard for Elira?” “Have you ever felt the same way?” “How can Elira feel more comfortable?” “What could she say or do?” “How can others help Elira?” “What small things could they do?”} &
Writing and some drawing \\
\midrule
High School &
3 × 3 table (rows = 3 challenges, columns = Problem, Tips for Elira, Tips for Others). &
\textit{“What is difficult for Elira in this situation?” “How might this affect her physically and emotionally?” “What could help Elira feel more comfortable and confident?” “What could she say or do?” “How can others be supportive?” “What small things could they do?”} &
Written reflection and analysis  \\
\bottomrule
\end{tabular}
}
\label{tab:worksheet}
\end{table*}

Each table also received three illustrated challenge cards (8.5×11 inches), with medieval visuals and age-appropriate text . After the story, each child received an individual worksheet (11×17 inches) tailored to their developmental stage. To maintain continuity, each worksheet included miniature versions of the challenge cards.
\begin{itemize}
    \item Elementary worksheets prompted drawing-based responses such as: “Show how she feels,” “Show what helps her feel better,” and “Show how she explains her special heart.” Large blank spaces encouraged creative expression, and counselors supported children by transcribing their exact words onto sticky notes placed beside the kids’ drawings, preserving the children’s voice rather than relying on adult interpretation.
    \item Middle school worksheets followed the same three-challenge structure but used a 3 × 3 table format, with columns labeled “Problem,” “Tips for Elira,” and “Tips for Others.” Middle schoolers reflected on prompts like “Why do you think this is hard for Elira?” and “Have you ever felt the same way?” Children reflected through short writing and some drawing.
    \item High school worksheets used the same structure but with more complex prompts, such as “What is difficult for Elira in this situation?” and “How might this affect her physically and emotionally?” This layered scaffolding supported older participants in analyzing Elira’s experiences at greater depth, while preserving structural consistency across age groups (see Table~\ref{tab:worksheet}).
\end{itemize}

Although playful, the quest served two purposes. (1) Provide a structured yet flexible method for children to articulate needs and propose strategies, generating directly comparable artifacts across age groups; (2) maintain coherence across challenges while adapting the complexity of prompts, the activity applied scaffolding theory \cite{wood1976tutoring}, balancing support with independence. This enabled mutual learning \cite{simonsen2013handbook,guha2004mixing}, as children externalized coping strategies and insights through the fictional lens of Lady Elira, while the research team observed how developmental stages shaped their articulation of needs, strategies, and emotions.

\subsubsection{Reflection Activity 2 (5 minutes)} 
\begin{figure}[h]
  \centering
  \includegraphics[width=\linewidth]{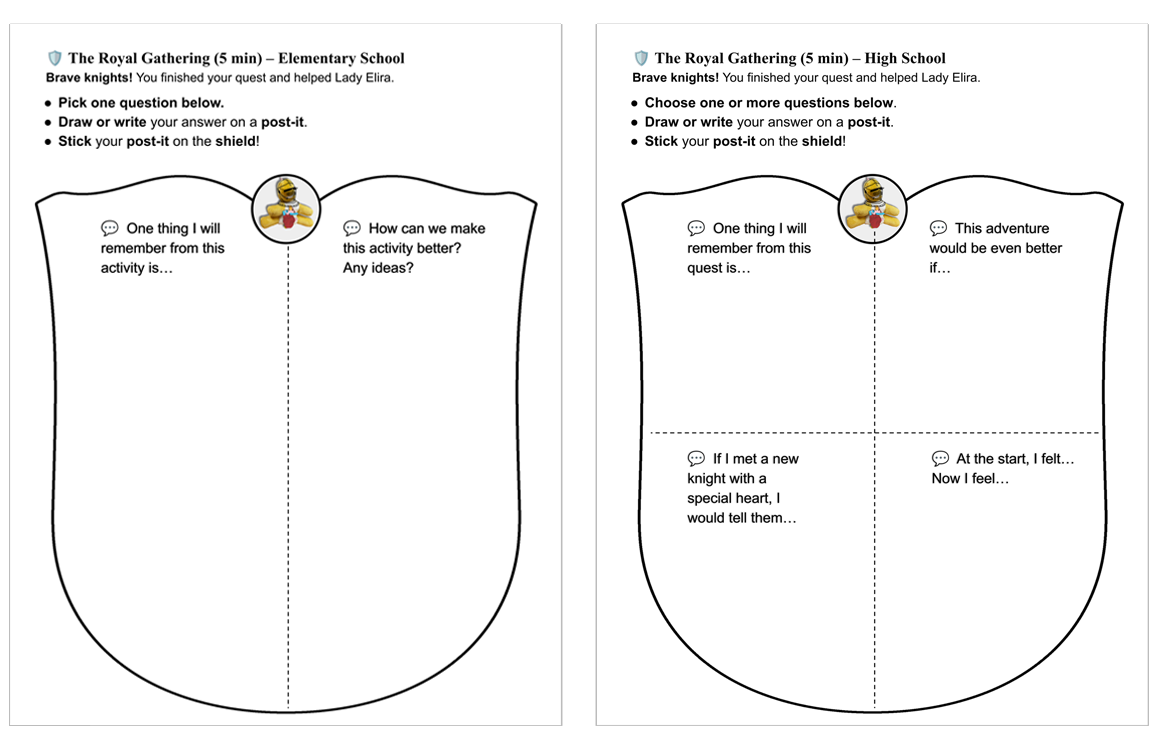}
  \caption{Reflection 2: Elementary school reflections (left) and high school reflections (right).}
  \Description{Two-panel figure showing Reflection 2 worksheets for elementary and high school groups. The left panel displays the elementary school worksheet, which uses a large shield outline divided into two halves by a vertical dotted line. At the top, children are instructed to choose one question, draw or write an answer on a sticky note, and place it on the shield. The left half of the shield includes a single prompt with a checkbox icon: “One thing I will remember from this activity is…”. The right half contains an open prompt: “How can we make this activity better? Any ideas?” The right panel shows the high school worksheet, also shaped like a shield but divided into four quadrants by vertical and horizontal dotted lines. At the top, students are instructed to choose one or more questions and respond on sticky notes. The upper-left quadrant repeats the prompt “One thing I will remember from this quest is…”. The upper-right quadrant reads “This adventure would be even better if…”. The lower-left quadrant asks “If I met a new knight with a special heart, I would tell them…”. The lower-right quadrant reads “At the start, I felt… Now I feel…”. Each shield includes a small circular image of the plush knight character at the top center.}
  \label{fig:Fig6}
\end{figure}

The workshop closed with a second reflection designed for closure and celebration. Children gathered around large, shield-shaped posters and added sticky notes (see Figure~\ref{fig:Fig6}). Elementary participants contributed to a two-section shield (“One thing I will remember\dots”; “How can we make this better?”). Middle and high school participants worked on a four-section shield (“At the start, I felt\dots Now I feel\dots”; “This adventure would be even better if\dots”; “One thing I will remember\dots”; “If I met a new knight with a special heart, I would tell them\dots”). The shared display supported both individual expression and collective recognition. Facilitators encouraged participation, provided writing help when needed, and validated contributions. To celebrate their role as co-designers, each child received a personalized “Guardian of Special Hearts” certificate and a colorful Wisdom Map Key summarizing heart anatomy. While not part of formal data collection, these tokens reinforced a positive ending and acknowledged children as active co-creators.

\begin{table*}[t]
\centering
\caption{Identified themes and subthemes.}
\resizebox{\textwidth}{!}{%
\begin{tabular}{{p{6cm} p{10cm}}}
\toprule
\textbf{Themes} & \textbf{Subthemes} \\
\midrule
Social Inclusion and Peer Participation
& \textbf{ES:} Overwhelmed by Feeling Different: Seeking Immediate Tangible Comforts \\
& \textbf{MS:} Fear of Peer Labeling: Pulling Back from Peers and Seeking Safe Spaces \\
& \textbf{HS:} Feeling Like a Burden: Finding Strategic Ways to Stay Involved \\
\midrule
Communicating About a ``Special Heart''
& \textbf{ES:} Limited Words: Emerging Narrative Through Symbols and Companions \\
& \textbf{MS:} Disclosure Hesitancy: Distancing Narrative Through Parents or Metaphors \\
& \textbf{HS:} Disclosure Risks: Owning Narrative Through Scripts and Control \\
\midrule
Medical Anxiety and Clinical Encounters
& \textbf{ES:} Strangeness of the Clinic: Finding Safety in Familiar Companions \\
& \textbf{MS:} Distrust in Providers: Depending on Parents and Tools for Mediation \\
& \textbf{HS:} Worry About Outcomes: Taking Agency in Care \\
\bottomrule
\end{tabular}
}
\label{tab:themes}
\end{table*}

These activities formed a layered design that moved from individual expression to collective reflection, balancing play with structured inquiry. The session aligned with \citet{sanders2014probes} typology of co-design tools: worksheets and posters served as generative toolkits, while plush knight character and Lady Elira acted as fictional props anchoring the narrative. To support
trust and comfortable sharing, each age group worked in small subgroups at separate tables with a familiar camp
counselor and a facilitator. We collected data from three sources: (1) children’s worksheets and posters, (2) audio and video recordings from tables, and (3) researcher field notes. For this paper, we analyzed the worksheets and posters. We reviewed field notes and recordings only for context and did not code or quote them; full analysis of the recordings is planned for future work.\enlargethispage*{16pt}

\subsection{Data Analysis}
We analyzed children’s worksheets and posters as the primary data source; their written responses and drawings on activity sheets were photographed and digitized for qualitative coding. For elementary schoolers, counselors transcribed children’s exact verbal explanations onto sticky notes attached to the drawings, and each drawing–note pair was treated as a unified analytic artifact to preserve younger children’s verbal descriptions and reduce misinterpretation. We imported all artifacts (including child’s written text, counselor-transcribed notes, and a brief researcher description of the drawing) into NVivo 15 and conducted thematic analysis \cite{braun2006ta,braun2021ta} to identify age-relevant patterns in how children with CHD navigate social participation, communication, and clinical encounters. 
The first author performed sentence-level open coding, preserving context (e.g., noting not only “fear” but also who felt it, in what situation, and why). Initial open codes were organized separately for each group—elementary school (ES), middle school (MS), and high school (HS)—and within each of the three predefined workshop challenge prompts (Social Exclusion and Difference, Communication Barriers, Medical Anxiety). This structure mirrored how the workshop elicited children’s input and ensured that age-based contrasts remained visible. This process generated 231 first-level codes. 

We then grouped semantically related first-level codes into broader mid-level codes by looking for ideas that appeared together across children’s worksheets. Most clustering was inductive, combining codes when children described similar experiences or reactions (e.g., different phrasings of feeling left out). In a smaller number of cases, we applied deductive consolidation when multiple codes clearly referred to the same underlying construct (e.g., several codes reflecting a child’s attempt to avoid embarrassment during medical conversations) \cite{bingham_data_2023, azungah_qualitative_2018}. This blended approach ensured that each mid-level code represented a coherent idea grounded in children’s words while reducing unnecessary fragmentation. Throughout, we wrote analytic memos to document emerging cross-age contrasts (i.e., concrete, immediate solutions in ES; identity threat, selective disclosure, and mortality awareness in MS; acceptance and boundary-setting in HS). This process resulted in 89 distinct mid-level codes.

The first author maintained the evolving codebook and provided weekly progress updates. During team meetings, we reviewed new codes, resolved ambiguities, and organized the mid-level codes into three analytic categories: Problems (challenges children encountered), Coping (child-initiated strategies), and Supports (others’ actions that aided the child). We used these three categories because they reflected the structure of the workshop activities and helped us see how children described the challenges they faced, how they responded, and what help they relied on. We then exported the three categorized sets of mid-level codes (Problems, Coping, Supports) to a virtual whiteboard (Miro) and used affinity clustering to group related codes. This step allowed us to keep codes within the appropriate category while comparing how ES, MS, and HS participants described similar challenges differently, ensuring that developmental differences remained visible and coherent across the dataset. Through this process, we identified the patterns that ultimately formed three high-level themes and nine subthemes (see Table~\ref{tab:themes}). These themes capture recurring developmental patterns across Problems, Coping, and Supports. Direct quotes were lightly edited for grammatical accuracy and clarity without altering meaning. The following section presents these findings in detail.
\enlargethispage*{16pt}

\begin{table*}[t]
\centering
\renewcommand{\arraystretch}{1.5} 
\caption{Developmental contrasts in coping strategies across CHD challenges.}
\resizebox{\textwidth}{!}{%
\begin{tabular}{p{3.5cm} p{4.5cm} p{4.5cm} p{4.5cm}}
\toprule
\textbf{Developmental Stage} & \textbf{Social Exclusion and Difference} & \textbf{Communication Barriers} & \textbf{Medical Anxiety} \\
\midrule
\raggedright\textbf{Elementary (ES)} &
Express overwhelming feelings; seek immediate comfort through breaks, toys, or small treats. &
Use simple words, drawings, or comfort objects as “speaking tools”; communication is blunt but expressive. &
Clinic feels strange and intimidating; cope through comfort objects, imagination, and parental presence. \\
\midrule
\raggedright\textbf{Middle school (MS)} &
Fear stigma and mortality; withdraw from activities; use subtle exit strategies; prefer safe/peer-like spaces. &
Hesitant; rely on parents, metaphors, or partial medical terms; create distance to avoid judgment. &
Distrust providers; depend on parents or tools; prefer step-by-step explanations and peer-like stories. \\
\midrule
\raggedright\textbf{High school (HS)} &
 Accept limits with responsibility; plan low-impact roles; use strategic inclusion to avoid burdening peers. &
 Prepare brief scripts and rehearsed disclosures; manage timing and narrative to stay in control. &
Worry about outcomes; prepare questions; seek direct engagement and agency in care. \\
\bottomrule
\end{tabular}
}
\label{tab:contrasts}
\end{table*}

\section{Findings}
The following findings address (RQ1) by detailing the challenges and coping strategies children identified at different developmental stages. We conducted a one-day co-design workshop consisting of four sessions, organized by age group: elementary school (ES), middle school (MS), and high school (HS). Across sessions, participants described how living with CHD shaped their experiences with peer participation, communication, and clinic encounters. Within these domains, challenges were paired with coping strategies and support from peers or adults. We present three key challenge areas (Social Exclusion and Difference, Communication Barriers, and Medical Anxiety) framed as problem-to-support narratives that emphasize developmental contrasts (see Table~\ref{tab:contrasts}).

\subsection{Social Inclusion and Peer Participation}
Children with CHD expressed a strong desire to remain part of peer groups, even when physical limitations made them feel different or “not normal,” especially during group activities like school sports. Coping strategies varied by age: elementary schoolers sought immediate comfort to manage feelings of overwhelm or exclusion; middle schoolers withdrew from activities to avoid being labeled;  high schoolers adapted their roles to avoid burdening others. While the desire for inclusion was consistent across age groups, the strategies reflected developmental differences, ranging from seeking immediate comfort and reassurance (ES) to retreating into safe spaces (MS) to role adjustment (HS).

\subsubsection{ES: Overwhelmed by Feeling Different: Seeking Immediate Tangible Comforts}
Elementary schoolers first noticed their differences during group play, realizing they could not keep up with peers. This awareness often led to confusion and immediate emotional responses, expressed through drawings, tears, asking to rest, or play. They managed these feelings through tangible comforts—taking breaks, sitting down, using plush toys, or having a small treat—to ease fatigue or sadness in the moment. For example, one participant drew a fictional peer with CHD crying and explained: 

    \begin{quote}
    \textit{``She is crying because she is sad about being left behind. Can’t run like them, but why? She needs to sit right there.''}
    \textit{(ES, Session 4, Cohort 9, P59)}
    \end{quote}
    
Rather than seeking long-term solutions, children turned to immediate comforts—taking a break, sitting down, or enjoying a treat like ice cream. As another participant shared, both in words and drawing: 

    \begin{quote}
    \textit{``My heart is special. Tired, let others know when I need a break. Maybe I’ll get some ice cream, I know it helps me.''}
    \textit{(ES, Session 4, Cohort 9, P54)}
    \end{quote}

Eager to stay involved, some children continued playing despite fatigue or pain, unaware of their limits. In such cases, trusted adults often intervened. Breaks, plush toys, or small treats helped them recover and rejoin play. Their sense of belonging was shaped more by immediate feelings of fatigue or confusion than by peer perception.

\subsubsection{MS: Fear of Peer Labeling: Pulling Back from Peers and Seeking Safe Spaces}
Middle schoolers, navigating early adolescence, felt increasingly self-conscious about their condition. Fearing judgment as “the sick kid,” worried about health risks, and were embarrassed by their differences, they avoided activities, relied on subtle exit strategies, and sought safe spaces where they did not need to explain themselves. Unlike elementary schoolers, who responded immediately to discomfort, they anticipated social risks and withdrew to protect themselves from labeling.

    \begin{quote}
    \textit{``It’s hard. I don’t want to go to school. I don’t want to be with other kids to tell them what’s wrong with me. I want to play, but if I don’t, they ask me why. And if I play, I think I’m going to die. Am I really gonna die?''}
    \textit{(MS, Session 3, Cohort 6, P40)}
    \end{quote}
    
Building on their growing awareness of their differences, middle schoolers also feared mortality, thinking minor activities might be dangerous. This fear made them hesitant to share their condition, worried it would cause concern and discourage group participation or play. As one child put it: 

    \begin{quote}
    \textit{``If I say it, then everyone thinks I’m dying. And when I don’t come to school, they think maybe I’m dead. I don’t want it, I don’t want to play. It’s hard.''}
    \textit{(MS, Session 3, Cohort 7, P43)}
    \end{quote}
    
To cope, they often withdraw from activities or use subtle “exit strategies”, like a pretend cough or a long water break, to rest without drawing attention to their condition. As P32 explained:

    \begin{quote}
    \textit{``Sometimes I cough to rest, or I get my water bottle. Sometimes I say, ‘ow, my knee.’ I do what other kids do, then it looks normal. Then they don’t ask, and don’t have to tell them.''}
    \textit{(MS, Session 2, Cohort 5, P32)}
    \end{quote}
    
In addition to withdrawing from group activities, they also sought places where they did not have to explain their condition, such as safe spaces like CHD-focused camps. Some chose friends with similar constraints; for instance, P42 said, \textit{``I like playing with Janet; she has asthma, so she can’t run as fast as I can :)''}. Beyond safe spaces, participants worried their condition might seem frightening to others.  To ease these fears, they relied on trusted adults, friends, or supportive tools that made explanations simpler and less intimidating.

Middle school children navigated difference through reassurance—avoiding activities, seeking safe spaces, and relying on trusted peers. This placed them between the immediate, emotion-driven responses of elementary schoolers and the planned strategies seen in high school.

\subsubsection{HS: Feeling Like a Burden: Finding Strategic Ways to Stay Involved}
High schoolers no longer felt embarrassed about their condition but worried about burdening peers or slowing group activities. To stay involved, they planned roles in advance, communicated their needs clearly, and prepared straightforward explanations. Their strategies contrasted with the avoidance seen in middle schoolers and the moment-to-moment reactions of elementary children, reflecting growing independence and self-advocacy. One participant, for example, opted to be a scorekeeper rather than a physically demanding player during a basketball game:

    \begin{quote}
    \textit{``I can’t always keep up, and that used to be super embarrassing. 
    Now I plan for it. I pick things like scorekeeping or calling subs, 
    so I’m still part of it, and I’m not making people wait for me.''}
    \textit{(HS, Session 1, Cohort 2, P11)}
    \end{quote}

In addition to choosing less physically demanding roles, participants communicated their needs in advance, allowing others to prepare without drawing attention. They valued when peers treated accommodations as routine rather than showing excessive sympathy. For example, P19 mentioned before an activity that they might need a short break and asked a friend to cover if needed:

    \begin{quote}
    \textit{``Usually, before we start, I’m like, ‘I might need a quick break. If I do the hand-heart, can you cover me?’ When my friends are chill and don’t make it a big deal, it’s easy. But I use simple explanations and a clear plan ahead so that I'm not scrambling.''}
    \textit{(HS, Session 1, Cohort 3, P19)}
    \end{quote}

Together, these strategies show how high schoolers balanced inclusion with responsibility by planning ahead and choosing roles that supported both their needs and their peer groups.

This theme illustrates how coping strategies evolve with age. Elementary schoolers expressed emotions openly and relied on immediate comforts; middle schoolers withdrew to avoid labeling; and high schoolers planned and adjusted their roles to stay involved. Across all ages, the desire to stay included remained consistent, but the strategies reflected each group’s developmental stage.

\subsection{Communicating About a “Special Heart”}
While the previous theme addressed inclusion, this theme focuses on how children chose to communicate about their “special heart.” Elementary schoolers used simple language or comfort objects to avoid feeling overwhelmed, middle schoolers relied on parents or metaphors to prevent judgment, and high schoolers prepared brief scripts to avoid complex explanations. Across age groups, the need to explain their condition remained constant, but their strategies varied by developmental stage, CHD understanding, and emotional readiness—whether relying on simple symbols and comfort objects (ES), metaphors and adult support (MS), or brief, practiced scripts (HS).

\subsubsection{ES: Limited Words: Emerging Narrative Through Symbols and Companions}
Elementary schoolers often relied on simple symbols or objects to communicate, rather than detailed verbal explanations. Though eager to stay involved—sometimes pushing through fatigue—they used concrete aids like drawings to bridge the gap. For example, one child placed a “tired” sticker next to a drawing of a weary figure and wrote:

    \begin{quote}
    \textit{``Too tired to play. I want to rest. Wait. My heart is ‘special’. Tired, let others know when I need a break.''}
    \textit{(ES, Session 4, Cohort 9, P54)}
    \end{quote}

They also used comfort objects, like stuffed animals, not only for emotional support but also as indirect tools for sharing feelings. One child described hugging their bear and speaking to it when sad or scared, which helped them open up to new friends in a gentle, non-verbal way: 

    \begin{quote}
    \textit{``Feels sad, scared, okay, but comforted by my stuffie (bear), when I’m embarrassed and want to talk to him [my new friend].''}
    \textit{(ES, Session 4, Cohort 9, P55)}
    \end{quote}

During the workshop, many participants shared that a “like me” toy could “do the talking,” acting as a translator in addition to being a comfort object. This allowed them to turn fears into approachable explanations. Even older participants reflected that they would have benefited from a plush toy when younger, as it could redirect attention and ease peer interactions. In this role, the plush acted as a translator, helping children express themselves when words were insufficient.

    \begin{quote}
    \textit{``She [fictional peer with CHD] feels scared. She could explain using pictures or a [character]; it’s like me, as it can do the talking for me. I can play with them and make friends.''}
    \textit{(ES, Session 4, Cohort 11, P67)}
    \end{quote}

In addition to these aids, children used symbolic drawings to express emotions they couldn’t verbalize. For example, P59 drew a broken heart and three stick figures, with one left out. Through their description, facilitators noted how these drawings conveyed confusion and embarrassment with complex language:

    \begin{quote}
    \textit{``She [fictional peer with CHD] feels confused and embarrassed when talking about her heart. She is scared they’re gonna laugh at her, she’s not like them.''}
    \textit{(ES, Session 4, Cohort 9, P59)}
    \end{quote}

Elementary participants communicated in immediate, symbolic ways that emerged through actions and objects rather than planned words. Their explanations emphasized expression rather than strategy—a pattern that becomes more structured in later age groups.

\subsubsection{MS: Disclosure Hesitancy: Distancing the Narrative Through Parents or Metaphors}
For middle schoolers, communicating with peers caused hesitation and fear. Middle schooler were often uncertain about how to explain their condition. They feared saying the wrong thing or giving an explanation that might make peers think they were seriously ill or fundamentally different, potentially leading to exclusion. As P40 expressed:

    \begin{quote}
    \textit{``Can you explain it? It’s hard. I don’t really want to say, ‘Hey, I’m different.’ Like, why should? If someone asks, I’ll tell them, but I don’t know what to say. But, I’m not just gonna walk up and be like, ‘Hey, I’m not like you.’ What if they stop being my friend or feel I'm dying? Then I’d have to make new friends all over again and explain again :(.''}
    \textit{(MS, Session 3, Cohort 6, P40)}
    \end{quote}

Building on this hesitation, middle schoolers often used fragments of biomedical language they heard from providers, blending clinical terms with child-friendly reasoning. Some described their heart as “not moving blood,” having “two holes,” or needing “more air.” While these phrases sometimes lent credibility, they also increased anxiety about misspeaking and often confused peers. To avoid errors, many preferred that parents or teachers explain on their behalf. As P39 shared:

    \begin{quote}
    \textit{``How to tell them oxygenated blood is not going to my heart wall. True? Not sure if my doc said or not. Ask parents, yes, they always talk for me and tell all my friends, I'm not dying. When they’re not there, I don't know how to say.''}
    \textit{(MS, Session 3, Cohort 6, P39)}
    \end{quote}

To manage uncertainty, some middle schoolers created emotional distance by using metaphors, fictional characters, or objects, making the explanation feel less personal. A common strategy was to compare their condition to a familiar character, transforming a potentially scary topic into something more relatable. This approach shifted attention away from themselves. As one participant explained, comparing their condition to Iron Man’s chest light:

    \begin{quote}
    \textit{``It’s hard to explain, so I just remarked it’s like Iron Man, Tony Stark’s chest light or something like that. It’s like Iron Man’s heart, you know? It helps me, but it’s different. People understand that better.''}
    \textit{(MS, Session 2, Cohort 4, P23)}
    \end{quote}

Besides metaphors, some wished for tools like a recognizable character to serve as conversation bridges. They hoped the “like-me, but clever” toy could help them navigate challenging aspects of explanations, turning something scary into something approachable, and make it easier to share and connect:

    \begin{quote}
    \textit{``If she [fictional peer with CHD] learns to say it [CHD explanation] so people see it’s not a big deal, she’ll feel better talking about it. The video we watched today [video character] wasn’t scary, and the yellow dino has a heart like mine but ‘clever’. Not scary, LOVE it. How cool, I can think of my heart like a game health bar that runs out faster than others.''}
    \textit{(MS, Session 3, Cohort 8, P49)}
    \end{quote}

These strategies show how middle schoolers managed disclosure by creating distance—using metaphors, characters, or adults as intermediaries rather than speaking directly. Their explanations were cautious and indirect, reflecting a transitional stage between the spontaneous expression seen in elementary school and the more deliberate communication that appears in high school.

\subsubsection{HS: Disclosure Risks: Owning the Narrative Through Scripts and Control}
By high school, participants actively managed how they disclosed their condition, treating communication as something to plan and control. While they did not express shame about having CHD, they remained cautious, aware that disclosure could shift peer dynamics, leading to overprotection or exclusion. To maintain agency, they carefully chose when, how, and what to share. As P12 explained, they developed short, matter-of-fact scripts that conveyed essential information without inviting pity or fear. They also invested time in learning about their condition so they could answer questions confidently and minimize misunderstandings:

    \begin{quote}
    \textit{``I just say, ‘I’ve got a heart thing, so I might sit out sometimes.’ People get it, and we move on. I don’t want a big reaction. I’ve learned to follow short ‘scripts’ and say it matter-of-factly, and it’s working. I’m slowly learning more about my heart so I can take control.''}
    \textit{(HS, Session 1, Cohort 2, P12)}
    \end{quote}

To ease communication, many rehearsed their phrasing in advance, ensuring disclosures felt smooth and controlled. Without preparation, they feared freezing, scrambling, or offering lengthy explanations that confused peers or invited unwanted questions. Their goal was to share just enough for peers to understand and move on. As one explained:

    \begin{quote}
    \textit{``I plan what to say so I don’t freeze. A quick explanation works better than all the details. I keep it short and clear, and not set off a flood of questions.''}
    \textit{(HS, Session 2, Cohort 3, P18)}
    \end{quote}

Practicing beforehand helped participants avoid “big deal” reactions from peers. Participants noted that excessive detail could confuse peers or make disclosure feel burdensome. They preferred brief, balanced explanations that were “fun but serious,” maintaining credibility without appearing fragile. To achieve this, some participants chose nonverbal tools—short videos, apps, or games—to make explanations easier and less personal. As one teen described:

    \begin{quote}
    \textit{``It's way easier if we just watch something that’s ‘official but fun official,’ like this [video character]. It has everything without me having to say it all. They get it and know it’s serious, but it doesn’t feel super heavy, and I can skip that ‘feel sorry for me in their eyes’. When I was a kid, I really wanted this funny [plush character], it says stuff to everyone without being about me. It’s clever and kind of knows us better than we do.''}
    \textit{(HS, Session 2, Cohort 3, P18)}
    \end{quote}

This balance—neither too intimidating nor too trivial—highlighted the maturity of their approach. For high schoolers, communication was about maintaining control, not hiding: choosing the timing, words, and details so they stayed in charge of their story.

This theme shows how children described their “special heart” in ways peers could understand without confusion or judgment. Strategies varied by age: elementary schoolers used symbols, simple words, and comfort objects as “speaking tools;” middle schoolers showed hesitation,  relying on parents or metaphors to prevent miscommunication; high schoolers prioritized independence, using brief scripts to avoid complicated explanations. Across all ages, the need to explain remained constant, but the methods reflected each group’s developmental stage, understanding of CHD, and ability to express emotions.

\subsection {Medical Anxiety and Clinical Encounters}
While earlier themes addressed inclusion and communication, this section explores how children experience and cope with anxiety during medical visits across developmental stages. Elementary schoolers found the clinic unfamiliar, middle schoolers doubted providers, and high schoolers focused on outcomes. Though fear and discomfort were common across all ages, coping strategies reflected each group’s developmental stage and relationship with providers—relying on comfort objects to reduce fear (ES), leaning on parents and supportive tools for reassurance (MS), or preparing questions and seeking direct engagement (HS).

\subsubsection{ES: Strangeness of the Clinic: Finding Safety in Familiar Companions}
Elementary schoolers often found the clinic unfamiliar and intimidating, responding primarily to the immediate sensory and emotional experience. They perceived providers as “strangers” and focused on the unpredictability of the environment, like needles, machines, and routines. One child drew a fictional peer with CHD crying in a clinic chair and explained:

    \begin{quote}
    \textit{``She is scared of the doctor. He feels like a stranger. She doesn’t know what’s going to happen.''}
    \textit{(ES, Session 4, Cohort 9, P59)}
    \end{quote}

Besides finding the clinic intimidating, elementary schoolers often had difficulty putting their medical experiences into words. They relied on short, direct statements like “It hurts” or “I don’t like it,” and using symbolic drawings, such as dark scribbles, to express their feelings. To cope, they relied on comfort objects, like their stuffed animals or blankets, treating them as protective shields. They imagined that these objects could “take the shot first” or “speak to the doctor” on their behalf. P67 wrote beside a drawing of a child holding a bear:

    \begin{quote}
    \textit{``When I’m scared, I hold my bear. I tell him, and then it’s okay. I don’t want to talk to the doctor. My bear talks and sometimes does the tests before me, they’re strange and kind, they let him in.''}
    \textit{(ES, Session 4, Cohort 11, P67)}
    \end{quote}

Elementary participants rarely suggested changes to communication with providers, focusing instead on emotional safety—staying close to parents, cuddling toys, or distracting themselves by drawing. For them, reassurance came from familiar objects/companions, making the clinic feel more familiar. P56 explained:

    \begin{quote}
    \textit{``I feel ‘worried’ and ‘nervous.’ She [fictional peer with CHD] feels scared going to the doctor, but she’s a little less shy after she has her bag of stuffies and is watching TV or drawing during the visit. When I bring my stuffie, I feel safe. If I don’t have him, I don’t feel safe at all.''}
    \textit{(ES, Session 4, Cohort 9, P56)}
    \end{quote}

Elementary schoolers described medical anxiety as rooted in unfamiliarity and strong emotions. They coped through comfort and imagination (e.g., familiar companions, predictable routines, trusted objects, and parental presence) rather than through the explanations/strategies that emerge in later stages.

\subsubsection{MS: Distrust in Providers: Depending on Parents and Tools for Mediation}
Middle schoolers found clinic visits frustrating and struggled to trust providers. They quickly noticed when medical explanations didn’t align with their experiences, leading to doubt and feelings of being misled. Many recalled false assurances, such as being told a procedure would not hurt, undermining their trust. One participant described:

    \begin{quote}
    \textit{``They say, ‘It won’t hurt,’ but it does. Then I don’t believe them anymore. They say things to keep you from being scared, but it’s not true. I hate being there.''}
    \textit{(MS, Session 3, Cohort 6, P41)}
    \end{quote}

Facing mistrust, middle schoolers often relied on parents to interpret or manage provider communication. This dependence often led to further confusion and anxiety, such as misunderstanding medical jargon or withholding information about upcoming visits. As P38 shared:

    \begin{quote}
    \textit{``My mom says what doc says, but I don’t think she gets it either. She just says the big words again, and I still don’t get it and get more scared. Sometimes she doesn’t tell me we’re going to the clinic, and then we’re just there. I don’t trust them.''}
    \textit{(MS, Session 2, Cohort 7, P38)}
    \end{quote}

To handle this mistrust and dependence, middle schoolers often withdrew or distracted themselves by staying quiet, avoiding eye contact, or focusing on a toy or technology during visits. They desired communication that matched their level: direct questions, slower pace, and step-by-step information. As P44 suggested:

    \begin{quote}
    \textit{``If doctors asked me, like, ‘are you okay?’ or just one question at a time, I’d talk more. Not all at once. Little by little, so I can answer.''}
    \textit{(MS, Session 3, Cohort 8, P44)}
    \end{quote}

Alongside their desire for more direct and paced communication, unlike elementary schoolers, they suggested tools to make medical conversations less intimidating. They found short videos, simple diagrams, and characters/toys to be effective conversation starters. Additionally, hearing stories about peers in similar situations helped them see positive outcomes as achievable. For example, P43 explained:

    \begin{quote}
    \textit{``Docs can help [Fictional girl with CHD] by telling her stuff like it’s ok to be nervous. If they have something like [plush toy], I liked it so much, it's not scary and is fun. Telling stories of other kids like [her] helps too.''}
    \textit{(MS, Session 3, Cohort 7, P43)}
    \end{quote}

Middle schoolers were in a state of dependent mistrust; they relied on adults to navigate clinical interactions but preferred that providers speak directly to them in simple language, take things step-by-step, and use supportive tools and peer stories. In contrast, elementary schoolers were less concerned with trust or outcomes and more focused on the unfamiliarity of the clinic, managing their anxiety through imagination, comfort objects, and parental support.

\subsubsection{HS: Worry About Outcomes: Taking Agency in Care}
By high school, participants approached medical visits with greater confidence, focusing on understanding their health outcomes. They recalled earlier confusion caused by unfamiliar providers and complex terminology.  Over time, they learned to trust their providers. This trust, however, did not eliminate anxiety. Their concerns shifted from procedures to outcomes—worrying about test results, potential surgeries, or worsening heart. As one teen explained:

    \begin{quote}
    \textit{``I don’t freak about needles anymore. I get nervous before appointments. Not because of shots, because of what they’ll say. Like, is it good news or bad news? I just want to know straight. I just want to know, what does it mean? If my numbers are bad, does that mean surgery? That’s what keeps me up.''}
    \textit{(HS, Session 1, Cohort 3, P21)}
    \end{quote}

As concerns shift towards outcomes, high schoolers also sought more independence. They felt frustrated when providers spoke only to their parents, making them feel invisible in their own care. They asked providers to speak directly to them, offer clear explanations, and acknowledge what they already understood. As P12 expressed:

    \begin{quote}
    \textit{``Talk to me, not just my mom. I can tell when I’m short of breath. When they don’t, it feels like I don’t exist. When I was little, I didn’t know much, and doctors were talking ‘big words’; we joked around. I really wanted them to talk to me in a way I could get. But now, I’ve learned, and I know more than my parents do. If you see a little kid, just talk to them in a way they get, without all the ‘big words’. They can handle it just fine without needing their parents around.''}
    \textit{(HS, Session 1, Cohort 2, P12)}
    \end{quote}

To manage their anxiety, high schoolers used self-management strategies similar to those from social inclusion. Many wrote down questions to remember during visits. Others practiced brief answers to routine prompts, like rating their pain or fatigue, to feel more in control. They preferred straightforward explanations over long lectures so they could process information without feeling overwhelmed. They also wanted the opportunity to lead the conversation, hearing the basics first and then choosing which details to explore:

    \begin{quote}
    \textit{``I don’t want a long lecture. Just tell me the facts, and I can ask more if I want. Tell me only what I need to know; then I’ll ask what I want to know more about!''}
    \textit{(HS, Session 2, Cohort 7, P16)}
    \end{quote}

In addition to self-preparation strategies, they valued simple tools that made medical conversations easier to understand. Diagrams, short videos, or characters/toys helped structure explanations without feeling overwhelmed. Importantly, they wanted these tools to act as bridges that supported their own voice rather than replacing it. As one teen explained:

    \begin{quote}
    \textit{``Hey, I think videos and apps can be really useful, and honestly, I still like having plush toys even at my age. But we need to talk. Don’t just rely on those tools; use them to make our conversations better, learn from them. As we’ve gotten older, we’ve started to really get to know our care team, and I want to share what I’ve learned and how I’m feeling. So, please talk to me directly, not just to my parents.''}
    \textit{(HS, Session 2, Cohort 4, P21)}
    \end{quote}

High schoolers demonstrated how anxiety shifted from fear of procedures to concern about outcomes. They managed this uncertainty through independence, preparation, and direct engagement with providers.

This theme explores how participants’ coping strategies shift over time to manage fear and build trust. Elementary schoolers relied on comfort objects and imagination to ease fears and make the clinic less intimidating. Middle schoolers sought reassurance through parents, tools, or peer-like stories to avoid feeling dismissed or misled. High schoolers managed anxiety by preparing questions and engaging providers directly. Across all age groups, children wanted to feel safe and understood, but the coping strategies they relied on reflected their developmental stage and evolving relationship with providers.

\section{Discussion}
Our cross-age co-design approach revealed developmental contrasts that single-age sessions rarely surface, answering (RQ1) by showing how children’s challenges and coping strategies change across stages, and (RQ2) by identifying design implications for adaptive interventions. Across all ages, children navigated two intersecting needs—belonging and communication—yet these needs manifested differently at each developmental stage. In our findings, belonging described how children negotiated inclusion and stigma, while communication captured the symbolic and strategic ways they shared or withheld information. In this section, we discuss design directions grounded in these developmental differences and suggest technology design strategies for interventions that grow alongside children.

\subsection{Design for Social Belonging Across Development}
Elementary school children described a “first shock of difference” when entering school without their parents, realizing their bodies were not the same as their friends. They coped by relying on familiar objects such as plush toys or shared activities like drawing, which provided comfort and opened conversations with other children without CHD \cite{barbazi2025boundary}. Trusted teachers also helped translate unfamiliar experiences. Foundational HCI research shows that tangible artifacts and shared materials support participation and make complex ideas accessible in early childhood \cite{walsh2010layered,guha2004mixing,druin1999cooperative}, while HCI healthcare research demonstrates how interactive toys and hospital play help young children regulate emotions and communicate \cite{isbister2021robot,lim2019distract,jeong2018huggable,slovak2016scaffolding, barbazi_understanding_2025}. Our findings extend this work by showing that for younger children, belonging is fundamentally rooted in comfort-seeking and tangible scaffolds that help normalize early experiences of difference \cite{alexandridis2021rubys}.

By middle school, awareness sharpened and was often coupled with fears of fragility, stigma, and mortality. Many withdrew from activities, worried peers would see them as fragile or assume the worst if they missed school. Some managed anxiety by stepping away from games, coughing to excuse themselves, or avoiding disclosure. Belonging shifted toward risk-management, aligning with research on health concealment to maintain normalcy \cite{hong2020diaries,mccarthy2017stigma,liu2015friends}. Design work shows that poorly framed support systems can unintentionally reinforce stigma \cite{su2024stigma}, while stigma-sensitive tools offer safer ways to negotiate disclosure \cite{su2024stigma,su2024cysticfibrosis}. Middle schoolers expressed a desire for “others like me,” echoing narrative-based scaffolding techniques that help youth externalize sensitive experiences \cite{warren_lessons_2022}. Similarly, clinical mentorship programs like iPeer2Peer \cite{stinson2016ipeer2peer} demonstrate that peer normalization is most effective when offered through age-matched peers.

High schoolers described accepting their differences yet managing the burden they might place on others—choosing low-impact roles, pacing themselves to avoid slowing peers, and balancing independence with self-protection.
Belonging becomes responsibility, aligning with research showing adolescents seeking contributive and leadership roles in care and peer networks \cite{wyche2025diabetes,saksono2020storywell}. 
Persona-based work similarly shows older youth co-construct guidance frameworks used by younger children \cite{warnestal_effects_2017}. This resonates with HCI work showing belonging sustained not only through participation but through leadership and guided responsibility across family care \cite{cha2023diabetes}, peer networks \cite{karusala2021courage}, and emerging online counseling platforms \cite{wang2025peer}. Our findings show that belonging in adolescence becomes tied to responsibility: youth maintain inclusion not only by “fitting in,” but by actively contributing to others’ well-being.

Synthesized across stages, these patterns outline a clear developmental trajectory: (1) ES = comfort-seeking through tangible anchors; (2) MS = risk-management through selective withdrawal and controlled disclosure; (3) HS = responsibility through contributive roles. Designing for belonging therefore requires systems that adjust to these shifting motivations rather than assuming a stable user identity.
This perspective also clarifies why prior peer-based interventions \cite{price2024painlogging,su2024stigma,su2024cysticfibrosis} have struggled to scale developmentally: they target one stage rather than supporting transitions across stages.
Based on the developmental differences we identified, multi-age child-to-child systems could normalize the initial shock for younger children, reduce stigma for middle schoolers, and provide leadership opportunities for adolescents. This aligns with research showing increased connectedness and resilience when adolescents mentor peers \cite{wang2025peer,berkanish2022peer,stinson2016ipeer2peer}.

Prior HCI research explored peer-based interventions—such as collaborative disclosure platforms \cite{price2024painlogging,su2024stigma,su2024cysticfibrosis} and family–peer support systems in chronic illness \cite{cha2023diabetes}—but these efforts typically target a single developmental stage. Our findings extend this work by showing how peer dynamics shift with age and by identifying belonging as an age-dependent pathway.

\subsection{Designing for Symptom Communication Across Childhood and Adolescence}
Elementary school children saw communication with friends and providers as interactions with “strangers”. They struggled to explain their “special heart”, relying on plush toys, drawings, and symbolic props to “speak for them” and rehearsed procedures using toys. Prior HCI work shows that tangible props and scaffolds support participation in challenging dialogues \cite{currin2021shy,jeong2018huggable,hiniker2017fictional} and that interactive materials ease fear in medical settings \cite{rashid2021pretendplay,vandelden2020spiroplay, barbazi_understanding_2025}. These symbolic companions are especially important when children first attempt disclosure without parental mediation, echoing findings that early childhood communication relies on external supports to bridge uncertainty \cite{jeong2018huggable}. Our findings highlight that these symbolic supports do not simply mediate fear—they provide the communicative structure children have not yet internalized.

Middle school children often withdrew or used metaphors (e.g., superheroes) and jargon to demonstrate sophisticated understanding while protecting vulnerability. They deferred to parents during clinical encounters and withheld details from peers to manage stigma. This aligns with research on stigma and disclosure management \cite{mccarthy2017stigma} and parental mediation in health communication \cite{pina2017family, hong2018odls}. Recent work shows that narrative tools can externalize private emotions without forcing direct disclosure \cite{warren_lessons_2022}, and our findings extend this by showing how metaphors become a strategic bridge between internal emotions and external communication demands.

High schoolers developed communicative scripts—preparing what to say to peers, simplifying explanations (avoiding jargon), and engaging providers directly. In our findings, scripts supported identity management, aligning with adolescent agency literature \cite{saksono2020storywell,hong2016care} and studies showing that communication rehearsal builds confidence and autonomy \cite{sonney2022asthmaapp,oygur2021wearables}. Persona-based work confirms that older youth can articulate structured communication patterns unavailable to younger children \cite{warnestal_effects_2017}.
Our findings therefore position communication as another developmental trajectory: (1) ES = symbolic aids; (2) MS = metaphor and controlled withholding; (3) HS = scripted agency.

Designing for symptom communication requires tools that reflect not only children’s cognitive abilities but also their motivational stance toward disclosure. Younger children need symbolic rehearsal; middle schoolers need protective metaphors and narrative shields; adolescents need tools that support agency and intentional self-presentation. These shifts offer concrete guidance for technology-mediated interventions that evolve with children with CHD rather than remaining static across ages.

\subsection{Toward Developmentally Adaptive Interventions} 
Across belonging and communication, a consistent pattern emerges: children’s strategies evolve systematically across ES–MS–HS. Prior pediatric HCI often highlights single-stage needs, but the cross-age contrasts surfaced here show why static, age-bounded technologies may fail to remain relevant as children grow. A tool effective at age seven (comfort-seeking, symbolic companions) may be misaligned at age thirteen (withdrawal, face-saving metaphors) and insufficient at seventeen (strategic scripts, agentic engagement). Just as pediatric cardiologists must repeatedly replace stents because most devices are built for static bodies rather than growing hearts, static technologies force children to outgrow their support systems instead of evolving with them. Designing for developmental trajectories ensures sustained relevance, enabling systems to flex with children’s evolving strategies for belonging, communication, and clinical trust. 

Our cross-age co-design workshops with the camp demonstrated one way to surface developmental contrasts without requiring longitudinal study. Holding diagnosis, environment, and facilitation constant allowed developmental differences—not contextual ones—to become visible. This comparative lens also reframes how we understand children’s participation: parents provide essential support, but older youth hold experiential knowledge that becomes developmentally meaningful when shared with younger children.  
These insights extend recent calls for pediatric systems that evolve across developmental periods \cite{warren_lessons_2022, warnestal_effects_2017, warren_codesign_2023} by showing how children’s strategies change and what technologies must adapt to. Together, these developmental patterns point to the need for cross-stage, adaptive interventions rather than static, age-siloed solutions—laying the foundation for systems that grow with children instead of requiring replacement at each developmental transition.

To illustrate these implications, we outline and example of how a single peer-to-peer system could adapt its design modalities across childhood. In early childhood, the system could function as a symbolic companion that uses simple stories contributed by older peers. In middle childhood, it could shift into narrative and metaphor-based interfaces that protect vulnerability and normalize fears. By adolescence, the system could unlock contribution layers that support mentoring, script-sharing, and leadership. A developmentally filtered repository would allow younger children to access resonant stories, middle schoolers to see evidence of thriving older peers, and adolescents to meaningfully support others. This approach turns peer knowledge into a living, child-generated resource that evolves with children instead of remaining static.

Finally, this contributes to HCI literature on mentorship systems \cite{karusala2021courage}, in-hospital peer systems \cite{haldar_patient_2020}, family-peer support \cite{cha2023diabetes}, and clinical peer interventions \cite{stinson2016ipeer2peer}, by showing how peer mechanisms can be structurally integrated into one adaptive system rather than delivered as disconnected, age-specific solutions. Addressing belonging, communication, and clinical anxiety as intertwined developmental pathways enables a unified approach to care. For example, adolescents could contribute disclosure scripts, middle schoolers could share metaphors or short videos that resonate with them, and younger children could engage with symbolic companions informed by older peers’ experiences. Participation across ages creates a living knowledge base, allowing the system itself to evolve as an iterative, peer-generated form of care. 
For chronic conditions, and especially for CHD, where adolescence as a lived stage is newly unfolding, this evolving peer-informed ecosystem offers children what previous generations never had: developmentally responsive support that grows with them into adulthood.

\section{Future Work And Limitations}
Our co-design workshops engaged children with CHD to derive design implications grounded in their developmental experiences. Although tailored to CHD, these implications may extend to other pediatric chronic conditions. As next steps, hands-on co-design with children should focus on prototyping technology-mediated systems that evolve with age—beginning in early childhood as comfort companions, growing into mediators that support communication and selective disclosure in middle school, and maturing into tools that scaffold agency and clinical engagement in adolescence—forming one developmentally adaptive system. 

While our age-comparative co-design, fictional framing, and camp-based setting created a supportive environment, they also introduce limitations. The camp provided a low-pressure space, but children may communicate differently in clinical or school contexts. The fictional storyworld—including a seven-year-old girl persona in a historical fantasy setting—reduced emotional risk but may have nudged ideas toward narrative or playful forms. Although we validated the persona’s content, we did not evaluate the persona’s performance as a facilitation mechanism or compare it to alternative scaffolding approaches. Warm-up activities helped reduce anxiety but may have shaped how children approached the main task. Because sessions ran from morning to midday, children’s energy—and facilitators’ energy and experience—may have varied across groups. Facilitators transcribed younger children’s words verbatim to reduce interpretation, yet minor phrasing influence remains possible. Finally, our cross-sectional design enabled same-day developmental comparison but cannot show how an individual child’s strategies evolve over time—a gap best addressed through longitudinal work, which remains challenging in pediatric specialty contexts.
\enlargethispage*{16pt}

\section{Conclusion}
Across four camp-based co-design workshops, we identified how children with CHD navigate inclusion, disclosure, and clinical encounters across developmental stages. Unlike studies treating children as a single group or using parents as proxies, we foregrounded developmental differences for design. The proposed interventions trace coping as an evolving trajectory: from comfort in early childhood, to mediated communication in middle school, to agency and clinical engagement in adolescence. Our discussion underscores the need for developmentally adaptive, child-centered participatory design grounded in lived experience. Although CHD-focused, these insights guide technology-mediated interventions for other chronic conditions, such as diabetes, as pediatric challenges shift with age. Beyond healthcare, developmentally adaptive approaches can benefit education and socio-emotional learning, where children’s coping, communication, and participation strategies evolve. By centering children’s voices, we advance pediatric systems that address current needs and grow alongside children’s developmental journeys.

\begin{acks}
This work was supported by the University of Minnesota Office of Discovery and Translation (ODAT) Pediatric Device Innovation Consortium (PDIC) grant. The authors are appreciative of the strong collaboration with the Department of Pediatrics at the University of Minnesota and Camp Odayin. We also thank former students Hannah Gootzeit, Irene Zeng, Grace Rubas, Jessica Jenkins Espinosa, Jonathan Jakubas, Levi Skelton, and Andy Thai for their invaluable contributions to this project.
\end{acks}

\bibliographystyle{ACM-Reference-Format}
\bibliography{main}

\end{document}